\documentclass[journal]{IEEEtran}

\ifCLASSINFOpdf
  \usepackage[pdftex]{graphicx}
  \DeclareGraphicsExtensions{.pdf,.jpeg,.png}
\else
  \usepackage[dvips]{graphicx}
\fi
\usepackage{epstopdf}
\usepackage[cmex10]{amsmath}
\usepackage{amssymb}
\usepackage{wasysym}
\usepackage{algorithm}
\usepackage{algorithmic}
\usepackage{array}
\usepackage{cite}
\usepackage{color}
\usepackage{url}

\usepackage{epsfig,latexsym}
\usepackage{flushend}
\usepackage{cite}
\usepackage{verbatim}
\usepackage{amsopn}
\usepackage{booktabs}

\usepackage{stfloats}
\usepackage{hyperref}
\usepackage{graphicx}
\usepackage{subfigure}

\begin{document}

\title{High-Resolution Channel Estimation for Frequency-Selective mmWave Massive MIMO System}

\author{Wenyan~Ma,~\IEEEmembership{Student~Member,~IEEE}, Chenhao~Qi,~\IEEEmembership{Senior~Member,~IEEE}, \\ and Geoffrey Ye Li,~\IEEEmembership{Fellow,~IEEE}
\thanks{This paper was presented in part at the IEEE ICC 2019, Shanghai, China.}
\thanks{This work was supported in part by National Natural Science Foundation of China under Grant 61871119, by Natural Science Foundation of Jiangsu Province under Grant BK20161428 and by the Fundamental Research Funds for the Central Universities. (\textit{Corresponding author: Chenhao~Qi})}
\thanks{Wenyan~Ma and Chenhao~Qi are with the School of Information Science and Engineering, Southeast University, Nanjing 210096, China (Email: qch@seu.edu.cn).}
\thanks{Geoffrey Ye Li is with the School of Electrical and Computer Engineering, Georgia Institute of Technology, Atlanta, GA, USA (Email: liye@ece.gatech.edu).}
}

\markboth{Submitted to IEEE Transactions on Wireless Communications}
{Shell \MakeLowercase{\textit{et al.}}: Bare Demo of IEEEtran.cls for Journals}

\maketitle

\begin{abstract}
In this paper, we develop two high-resolution channel estimation schemes based on the estimating signal parameters via the rotational invariance techniques (ESPRIT) method for frequency-selective millimeter wave (mmWave) massive MIMO systems. The first scheme is based on two-dimensional ESPRIT (TDE), which includes three stages of pilot transmission. This scheme first estimates the angles of arrival (AoA) and angles of departure (AoD) and then pairs the AoA and AoD. The other scheme reduces the pilot transmission from three stages to two stages and therefore reduces the pilot overhead. It is based on one-dimensional ESPRIT and minimum searching (EMS). It first estimates the AoD of each channel path and then searches the minimum from the identified mainlobe. To guarantee the robust channel estimation performance, we also develop a hybrid precoding and combining matrices design method so that the received signal power keeps almost the same for any AoA and AoD. Finally, we demonstrate that the proposed two schemes outperform the existing channel estimation schemes in terms of computational complexity and performance.

\end{abstract}
\begin{IEEEkeywords}
Millimeter wave communications, channel estimation, hybrid precoding, massive MIMO.
\end{IEEEkeywords}

\section{Introduction}
Millimeter wave (mmWave) communication is a promising technology for next generation wireless communications~\cite{heath2016overview,li20155G}. Different from most wireless communication systems operating at carrier frequencies below 6 GHz, mmWave communication systems make use of spectrum from 30 GHz to 300 GHz. The main benefits of using mmWave carrier frequencies are the abundant frequency spectrum resources and high data rates. To save the power consumed by radio-frequency (RF) chains and reduce cost, the hybrid structure is usually used, where a small number of RF chains are connected to a large antenna array. In order to form directional signal transmission with parallel data streams, the hybrid structure often includes the analog precoding and digital precoding~\cite{alkhateeb2014mimo,han2015large,choi2017resolution}.

Due to the massive antennas at both the transmitter and receiver in mmWave communications, the size of channel matrix is very large. Therefore, estimation of mmWave channels is usually time consuming. Moreover, the mmWave channels are frequency-selective in most of application environments~\cite{javier2018frequency,dai2016estimation,wang2018spatial,gonzalez2018channel,wang2018spatial2}, which brings more challenges for channel estimation.

Early work on mmWave massive MIMO channel estimation usually assumes that the channels are with flat-fading, which makes channel estimation much easier. There have been many channel estimation schemes  for the flat-fading mmWave systems. The estimating signal parameters via rotational invariance techniques (ESPRIT) method is the classical AoA estimation method used in radar systems. However, since the mmWave massive MIMO system with hybrid precoding and combining is totally different from the radar system, it is difficult to directly use the ESPRIT method. A scheme based on the ESPRIT and a scheme based on multiple signal classification (MUSIC) are proposed in~\cite{liao20172d} and~\cite{guo2017millimeter}, respectively. In order to eliminate the effect of the hybrid precoding and combining so that the ESPRIT and MUSIC methods can be directly used, these two schemes have to turn off approximately half of the antennas so that the number of active antennas is equal to that of time slots for channel estimation, which reduces the total transmission power and signal coverage. Two schemes based on the beamspace ESPRIT method are proposed in~\cite{zhang2017channel} and~\cite{zhang2017channel2}. However, The range of angles of arrival (AoA) and angles of departure (AoD) that can be estimated is only  approximately an eighth of the whole range of the AoA and AoD. Therefore these two schemes  have narrow  signal coverage and cannot estimate mmWave channels with any AoA and AoD. An identity matrix approximation (IA)-based channel estimation scheme has been also developed in~\cite{ma2018beamspace}, where mmWave channels are assumed to have only the line-of-sight (LOS) path.

Recent work demonstrates that the mmWave massive MIMO channels are actually frequency-selective. In~\cite{venugopal2017time}, a channel estimation scheme based on orthogonal matching pursuit (OMP) is proposed by exploring the sparsity of beamspace channels. A channel estimation scheme based on simultaneous weighted orthogonal matching pursuit (SWOMP) in~\cite{javier2018frequency} whitens the spatial noise components to improve the estimation accuracy using the OMP method. However, due to the limited beamspace resolution, the sparsity of beamspace channel may be impaired by power leakage~\cite{dai2017tras}, which brings extra challenges for the sparse recovery. To solve this problem, a distributed grid matching pursuit (DGMP)-based channel estimation scheme is developed to iteratively detect and adjust the channel support~\cite{dai2016estimation}.

In this paper, we investigate high-resolution channel estimation for frequency-selective mmWave massive MIMO systems, where orthogonal frequency division multiplexing (OFDM) is employed. We propose two channel estimation schemes based on ESPRIT~\cite{roy1989esprit} for mmWave massive MIMO channels with OFDM transmission. Different from the existing schemes based on the ESPRIT and MUSIC methods~\cite{liao20172d,guo2017millimeter}, our schemes can estimate frequency-selective channels and only need to turn off one antenna, which has little impact on the total transmission power. The contribution of this paper is summarized as follows.

1) We propose a two-dimensional ESPRIT (TDE)-based channel estimation scheme that includes three stages of pilot transmission. We use the first and second stages to estimate the AoA and use the first and third stages to estimate the AoD. We also develop an algorithm to pair the AoA and AoD.

2) To reduce the overhead of pilot transmission, we propose an one-dimensional ESPRIT and minimum searching (EMS)-based channel estimation scheme, which only requires two stages of pilot transmission. We estimate the AoD by first converting it into individual estimation of each channel path and then searching the minimum from the identified mainlobe.

3) We develop a hybrid precoding and combining matrices design method so that the received signal power keeps almost the same for any AoA and AoD to guarantee stable channel estimation performance. We consider the row-wise design of the hybrid combining matrix, where the least-square (LS) estimation with two undetermined variables is first obtained and then a power-ratio maximization criterion is used to determine two variables.

The rest of the paper is organized as follows. In Section II, we introduce the system model and formulate the problem of channel estimation for frequency-selective mmWave massive MIMO systems with hybrid precoding and combining. In Sections~III and IV, we propose two high-resolution channel estimation schemes. In Section~V, we develop a hybrid precoding and combining matrices design method. The simulation results are provided in Section VI. Finally, Section VII concludes the paper.

The notations are defined as follows. Symbols for matrices (upper case) and vectors (lower case) are in boldface. $(\cdot)^T $, $(\cdot)^H $, $(\cdot)^* $, $(\cdot)^{-1} $ and $(\cdot)^{\dag} $ denote the transpose, conjugate transpose (Hermitian), conjugate, inverse, and pseudo inverse, respectively. We use $\boldsymbol{I}_{L}$ and $\textbf{1}_{L}$ to represent identity matrix of size $L$ and vector of size $L$ with all entries being 1. The set of $M\times{N}$ complex-valued matrices is denoted as $\mathbb{C}^{M\times{N}}$. $\otimes$ and $\circ$ denote Kronecker product and Khatri-Rao product, respectively. We use $\textrm{vec}(\cdot)$ and $\textrm{diag}(\boldsymbol{a})$ to denote vectorization and the square diagonal matrix with the elements of vector $\boldsymbol{a}$ on the main diagonal. We use $\mathbb{E}\{\cdot\}$ to denote expectation. Order of complexity is denoted as $\mathcal{O}(\cdot)$. Zero vector of size $M$ is denoted as $\boldsymbol{0}^M$, while $M\times{N}$ zero matrix is denoted as $\boldsymbol{0}_{M\times N}$. $\|\cdot\|_2$ and $\|\cdot\|_F$ denote $l_2$-norm of a vector and Frobenius norm of a matrix, respectively. Entry of $\boldsymbol{A}$ at the $p$th row and $q$th column is denoted as $\boldsymbol{A}[p,q]$. $\langle\cdot\rangle$, $\textrm{Tr}(\cdot)$, $\cup$, $\cap$ and $\emptyset$ denote round function, trace, union operator, intersection operator and empty set, respectively. We use $\mathbb{Z}$ to denote set of integer. Complex Gaussian distribution is denoted as $\mathcal{CN}$.

\section{System Model and Problem Formulation}
After introducing OFDM based mmWave massive MIMO system, we then analyze the properties of mmWave channels and formulate the problem of channel estimation.
\subsection{System Model}
\begin{figure}[!t]
\centering
\includegraphics[width=85mm]{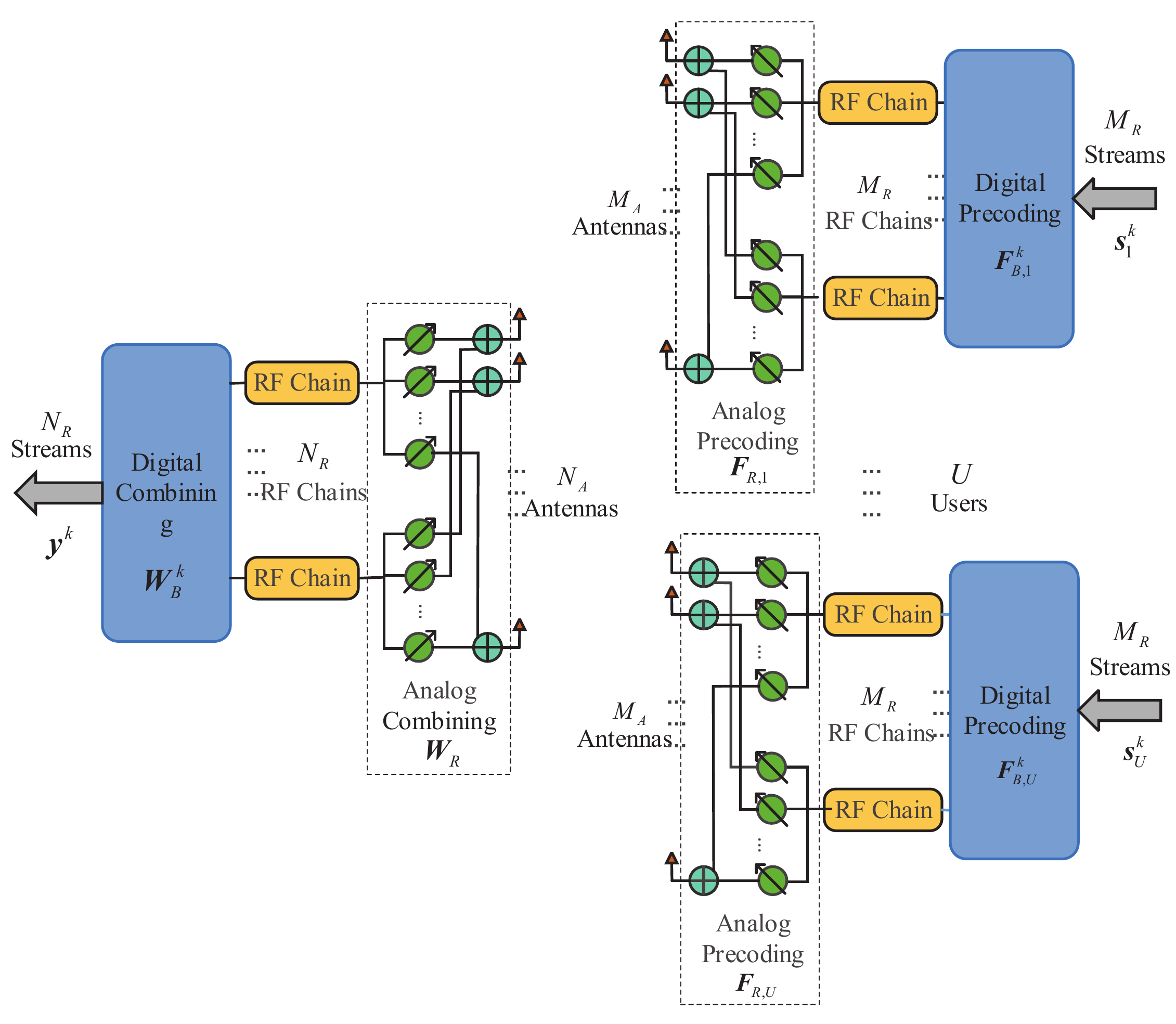}
\caption{Block diagram of uplink transmission in multi-user mmWave massive MIMO systems.}
\label{FIG8}
\end{figure}

We consider an uplink multi-user mmWave massive MIMO system comprising a base station (BS) and $U$ users as shown in Fig.~\ref{FIG8}. OFDM modulation with $K$ subcarriers is employed to deal with the frequency-selective fading channels. Both the BS and users are equipped with uniform linear arrays (ULAs). Let $N_A$, $M_A$, $N_R$ and $M_R$ denote the numbers of antennas at the BS and at each user and the numbers of RF chains at the BS and at each user, respectively. The mmWave massive MIMO system is with hybrid precoding and combining. Therefore, the number of RF chains is much smaller than that of antennas, i.e., $N_R \ll N_A$ and $M_R \ll M_A$~\cite{ma2017channel}.

For uplink transmission, each user performs analog precoding in RF and digital precoding in the baseband while the BS performs analog combining in RF and digital combining in the baseband~\cite{sun2019beam}. The received signal vector on the $k$th OFDM subcarrier by the BS can be represented as
\begin{equation}\label{UpSig}
\boldsymbol{y}^k = \boldsymbol{W}_B^k \boldsymbol{W}_R  \sum_{u=1}^{U} \boldsymbol{H}_u^k \boldsymbol{F}_{R,u} \boldsymbol{F}_{B,u}^k \boldsymbol{s}_u^k + \boldsymbol{W}_B^k \boldsymbol{W}_R \boldsymbol{n},
\end{equation}
where $\boldsymbol{F}_{B,u}^k\in{\mathbb{C}^{M_R\times{M_R}}}$, $\boldsymbol{F}_{R,u}\in{\mathbb{C}^{M_A\times{M_R}}}$, $\boldsymbol{W}_B^k\in{\mathbb{C}^{N_R\times{N_R}}}$, and $\boldsymbol{W}_R\in{\mathbb{C}^{N_R\times{N_A}}}$ are the digital precoding matrix, analog precoding matrix, digital combining matrix, and analog combining matrix for the $u$th user, respectively. To normalize the power of the hybrid precoder and combiner, we set $\| \boldsymbol{F}_{R,u} \boldsymbol{F}_{B,u}^k \| _F ^2= 1$ and $\| \boldsymbol{W}_B^k \boldsymbol{W}_R \| _F ^2= 1$. Note that the analog precoding and combining matrices are frequency-independent  while the digital precoding and combining matrices depend on subcarriers~\cite{javier2018frequency}. Denote $\boldsymbol{s}_u^k\in{\mathbb{C}^{M_R}}$ as the signal vector satisfying $\mathbb{E}\{\boldsymbol{s}_u^k{\boldsymbol{s}_u^k}^H\}=\boldsymbol{I}_{M_R}$, $\boldsymbol{n} \in{\mathbb{C}^{N_A}} $ as additive white Gaussian noise (AWGN) vector satisfying $\boldsymbol{n} \sim \mathcal{CN}(0,\sigma^{2}\boldsymbol{I}_{N_A})$, and $\boldsymbol{H}_u^k\in{\mathbb{C}^{N_A\times{M_A}}}$ as the channel matrix on the $k$th subcarrier between the BS and the $u$th user and can be expressed as~\cite{javier2018frequency}
\begin{equation}\label{Huk}
\boldsymbol{H}_{u}^k = \sum_{d=0}^{D-1} \boldsymbol{H}_{u,d}e^{-j\frac{2\pi kd}{K}},
\end{equation}
where $D$ denotes the number of delay taps of the channel.

According to the widely used Saleh-Valenzuela channel model~\cite{heath2016overview}, the channel matrix at the $d$th delay tap can be expressed as
\begin{equation}\label{ULAchannelmodel}
\boldsymbol{H}_{u,d}=\gamma\sum_{i=1}^{L_u} g_{u,i} p_{rc,u}(dT_s-\tau_{u,i}) \boldsymbol{\alpha} (N_A,\theta_{u,i}) \boldsymbol{\alpha}^{H} (M_A,\phi_{u,i}),
\end{equation}
where $L_{u}$, $g_{u,i}$, $p_{rc,u}(t)$, $T_s$, and $\tau_{u,i}$ denote the total number of resolvable paths, the channel gain, the pulse shaping, the sampling
interval, and the delay of the $i$th path for the $u$th user, respectively, $\gamma \triangleq \sqrt{N_AM_A / L_u}$, and the steering vector
\begin{equation}\label{alpha1}
\boldsymbol{\alpha}(N,\theta)=\frac{1}{\sqrt{N}}\left[1,e^{j\pi\theta},...,e^{j\pi\theta(N-1)}\right]^{T}.
\end{equation}
Denote the AoA and AoD of the $i$th path of the $u$th user as $\vartheta_{u,i}$ and $\varphi_{u,i}$, respectively, which are uniformly distributed over $[-\pi,\pi)$ \cite{alkhateeb2014channel,rappaport2015wideband}. Then in  (\ref{ULAchannelmodel}) $\theta_{u,i} \triangleq \sin{\vartheta_{u,i}}$ and $\phi_{u,i} \triangleq \sin{\varphi_{u,i}}$ if the distances between adjacent antennas at the BS and the users are with half-wave length.

\subsection{Analysis of Frequency-Selective mmWave Channel}
Denote $\widetilde{g}_{u,d,i} \triangleq g_{u,i} p_{rc,u}(dT_s-\tau_{u,i})$, $i=1,2,\ldots,L_u$. Then, from (\ref{ULAchannelmodel}), the channel matrix at the $d$th delay tap can be represented in a more compact form as
\begin{equation}
\boldsymbol{H}_{u,d}=\gamma \boldsymbol{A}_{u,R} \boldsymbol{\Delta}_{u,d} \boldsymbol{A}_{u,T}^H,
\end{equation}
where $\boldsymbol{A}_{u,R} \in{\mathbb{C}^{N_A \times L_u}}$, $\boldsymbol{A}_{u,T} \in{\mathbb{C}^{M_A \times L_u}}$, and $\boldsymbol{\Delta}_{u,d} \in{\mathbb{C}^{L_u \times L_u}}$ are denoted as
\begin{align}
\boldsymbol{A}_{u,R} &\triangleq [\boldsymbol{\alpha} (N_A,\theta_{u,1}),\boldsymbol{\alpha} (N_A,\theta_{u,2}),\ldots,\boldsymbol{\alpha} (N_A,\theta_{u,L_u})],  \notag \\
\boldsymbol{A}_{u,T} &\triangleq [\boldsymbol{\alpha} (M_A,\phi_{u,1}),\boldsymbol{\alpha} (M_A,\phi_{u,2}),\ldots,\boldsymbol{\alpha} (M_A,\phi_{u,L_u})],  \notag \\
\boldsymbol{\Delta}_{u,d} &\triangleq \textrm{diag}\left( [ \widetilde{g}_{u,d,1},\widetilde{g}_{u,d,2},\ldots,\widetilde{g}_{u,d,L_u} ] \right).
\end{align}
Then (\ref{Huk}) can be further rewritten as
\begin{align}\label{Huk2}
\boldsymbol{H}_{u}^k & =  \gamma \boldsymbol{A}_{u,R} \boldsymbol{\Lambda}_u^k \boldsymbol{A}_{u,T}^H,
\end{align}
where $\boldsymbol{\Lambda}_u^k \triangleq \sum_{d=0}^{D-1} \boldsymbol{\Delta}_{u,d} e^{-j\frac{2\pi kd}{K}} \in{\mathbb{C}^{L_u \times L_u}}$ is a diagonal matrix with $L_u$ nonzero entries. $\boldsymbol{\Lambda}_u^k[i,i] = \sum_{d=0}^{D-1} \widetilde{g}_{u,d,i} e^{-j\frac{2\pi kd}{K}}$ corresponds to the channel gain for the $i$th path at the $k$th subcarrier. From (\ref{Huk2}), all these $K$ subcarriers share the same AoA and AoD. Therefore we can estimate the AoA and AoD utilizing any subcarrier. Without loss of generality, we use the first subcarrier, i.e., $k=0$, to estimate the AoA and AoD. The mmWave channels are frequency-selectivity because all the subcarriers have different channel gains. After the AoA and AoD are estimated, the channels for all $K$ subcarriers are reconstructed using the pilot sequences transmitted at all $K$ subcarriers.


\subsection{Problem Formulation}

\begin{figure}[!t]
\centering
\includegraphics[width=85mm]{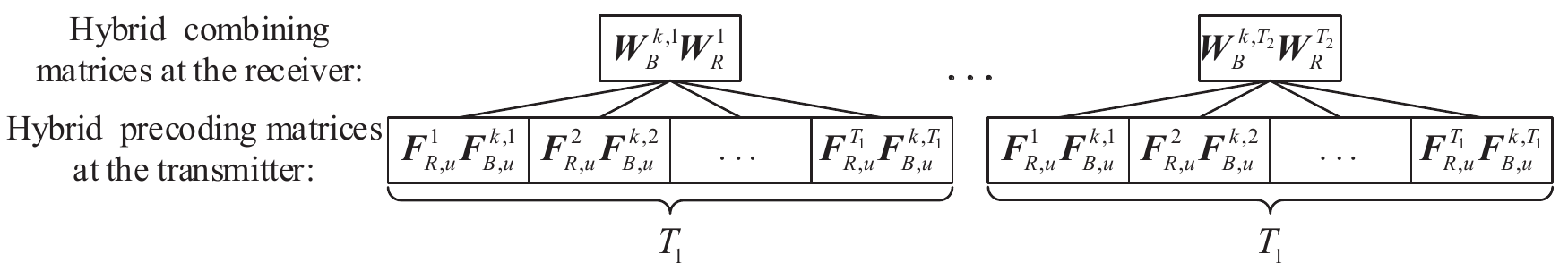}
\caption{$T_1 T_2$ repetitive transmission of pilot sequence for the $u$th user.}
\label{FIG9}
\end{figure}

Note that $\boldsymbol{y}^k$ in (\ref{UpSig}) is a combination of signal from different users. We use $T_1$ different digital precoding matrices and analog precoding matrices, denoted as $\boldsymbol{F}_{B,u}^{k,t_1} \in{\mathbb{C}^{M_R\times{M_R}}} $ and $\boldsymbol{F}_{R,u}^{t_1}\in{\mathbb{C}^{M_A\times{M_R}}}$, respectively at the $u$th user. We use $T_2$ different digital combining matrices and analog combining matrices, denoted as $\boldsymbol{W}_{B}^{k,t_2}\in{\mathbb{C}^{N_R\times{N_R}}}$ and $\boldsymbol{W}_{R}^{t_2}\in{\mathbb{C}^{N_R\times{N_A}}}$, respectively at the BS. The superscript $t_1$ and $t_2$ represent the $t_1$th precoding matrix and the $t_2$th combining matrix, respectively. To distinguish signals from different users at the BS, each user repeatedly transmits an orthogonal pilot sequence $\boldsymbol{p}_u^k \in{\mathbb{C}^U} $ for $T_1 T_2$ times. For simplicity, each user transmits the same pilot sequence for all $M_R$ RF chains, where the pilot matrix for the $u$th user can be expressed as $\boldsymbol{P}_u^k \triangleq [\boldsymbol{p}_u^k,\boldsymbol{p}_u^k,\ldots,\boldsymbol{p}_u^k]^H = \textbf{1}_{M_R}{\boldsymbol{p}_u^k}^H \in{\mathbb{C}^{M_R\times{U}}}$. The channel is assumed to be time-invariant during $T \triangleq T_1 T_2 U$ time slots.

It is worth pointing out that for mmWave communications, although the channel coherence time is usually small due to the high carrier frequency,
it still contains quite a large number of symbols thanks to the large mmWave bandwidth. For example, when the carrier frequency is 28 GHz and the bandwidth is 1 GHz, a maximum speed of 30 m/s results in the small channel coherence time of 0.36 ms. However, the symbol duration is in the order of 1 ns, which means that the small channel coherence time still contains 400,000 symbols~\cite{dai2017tras}.

During the $T_1$ repetitive transmission of pilot sequence from the $((t_2-1)T_1+1)$th transmission to $(t_2 T_1)$th transmission as shown in Fig.~\ref{FIG9}, we use $T_1$ different $\boldsymbol{F}_{B,u}^{k,t_1}$ and $\boldsymbol{F}_{R,u}^{t_1}$ for hybrid precoding while using the same $\boldsymbol{W}_{B}^{k,t_2}$ and $\boldsymbol{W}_{R}^{t_2}$ for hybrid combining. The received pilot matrix $\boldsymbol{Y}^{k,t_1,t_2}\in{\mathbb{C}^{N_R\times{U}}}$ can be denoted as
\begin{equation}
\boldsymbol{Y}^{k,t_1,t_2} = \boldsymbol{W}_B^{k,t_2} \boldsymbol{W}_R^{t_2}  \sum_{u=1}^{U} \boldsymbol{H}_u^k \boldsymbol{F}_{R,u}^{t_1} \boldsymbol{F}_{B,u}^{k,t_1} \boldsymbol{P}_u^k  +  \widetilde{\boldsymbol{N}}^{t_1,t_2},
\end{equation}
where $\widetilde{\boldsymbol{N}}^{t_1,t_2} \triangleq \boldsymbol{W}_B^{k,t_2} \boldsymbol{W}_R^{t_2} \boldsymbol{N}^{t_1,t_2} \in{\mathbb{C}^{N_R\times{U}}}$. Each entry of the AWGN matrix $\boldsymbol{N}^{t_1,t_2}\in{\mathbb{C}^{N_A\times{U}}}$ is with independent complex Gaussian distribution with zero mean and variance of $\sigma^2$. Due to the orthogonality of $\boldsymbol{p}_u^k$, i.e., ${\boldsymbol{p}_u^k}^H$$\boldsymbol{p}_u^k$=1 and ${\boldsymbol{p}_u^k}^H$$\boldsymbol{p}_i^k=0$, $\forall u,i\in \{1,2,\ldots,U\}$, $i\neq{u}$ \cite{he2020propagation}, we can obtain the measurement vector $\boldsymbol{r}_u^{k,t_1,t_2}\in{\mathbb{C}^{N_R}}$ for the $u$th user by multiplying $\boldsymbol{Y}^{k,t_1,t_2}$ with $\boldsymbol{p}_u^k$ as
\begin{equation}
\boldsymbol{r}_u^{k,t_1,t_2} \triangleq \boldsymbol{Y}^{k,t_1,t_2} \boldsymbol{p}_u^k = \boldsymbol{W}^{k,t_2} \boldsymbol{H}_u^k \boldsymbol{f}^{k,t_1}_u + \widetilde{\boldsymbol{n}}^{t_1,t_2},
\end{equation}
where
\begin{align}\label{Wt2}
\boldsymbol{W}^{k,t_2}  \triangleq \boldsymbol{W}_B^{k,t_2}\boldsymbol{W}_R^{t_2},&~~\boldsymbol{f}^{k,t_1}_u  \triangleq{\boldsymbol{F}_{R,u}^{t_1} \boldsymbol{F}_{B,u}^{k,t_1} \textbf{1}_{M_R}}, \notag \\ \widetilde{\boldsymbol{n}}^{t_1,t_2} \triangleq &\widetilde{\boldsymbol{N}}^{t_1,t_2} \boldsymbol{p}_u^k.
\end{align}
Denote $T_3 \triangleq T_2N_R$. We stack the $T_2$ received pilot sequences together and have
\begin{equation}
\boldsymbol{r}_u^{k,t_1} = \boldsymbol{W}^k \boldsymbol{H}_u^k \boldsymbol{f}^{k,t_1}_u + \widetilde{\boldsymbol{n}}^{t_1},
\end{equation}
where
\begin{equation}\label{DefinitionW}
\begin{split}
 \boldsymbol{r}_u^{k,t_1} &\triangleq [(\boldsymbol{r}_u^{k,t_1,1})^T , (\boldsymbol{r}_u^{k,t_1,2})^T , \ldots , (\boldsymbol{r}_u^{k,t_1,T_2})^T]^T \in{\mathbb{C}^{T_3}}, \\
 \boldsymbol{W}^k &\triangleq [(\boldsymbol{W}^{k,1})^T , (\boldsymbol{W}^{k,2})^T , \ldots , (\boldsymbol{W}^{k,T_2})^T]^T \in{\mathbb{C}^{T_3 \times{N_A}}}, \\
 \widetilde{\boldsymbol{n}}^{t_1} &\triangleq [(\widetilde{\boldsymbol{n}}^{t_1,1})^T , (\widetilde{\boldsymbol{n}}^{t_1,2})^T , \ldots , (\widetilde{\boldsymbol{n}}^{t_1,T_2})^T]^T \in{\mathbb{C}^{T_3}}.
\end{split}
\end{equation}
Note that $T_3$ is the row size of $\boldsymbol{W}^k$. Denote
\begin{equation}\label{DefinitionF}
\begin{split}
 \boldsymbol{R}_u^k &\triangleq [\boldsymbol{r}_u^{k,1} , \boldsymbol{r}_u^{k,2} , \ldots , \boldsymbol{r}_u^{k,T_1}]  \in{\mathbb{C}^{T_3 \times T_1}}, \\
 \boldsymbol{F}_u^k &\triangleq [\boldsymbol{f}^{k,1}_u , \boldsymbol{f}^{k,2}_u ,  \ldots , \boldsymbol{f}^{k,T_1}_u] \in{\mathbb{C}^{M_A \times T_1}}, \\
 \widetilde{\boldsymbol{n}} &\triangleq [\widetilde{\boldsymbol{n}}^{1} , \widetilde{\boldsymbol{n}}^{2} , \ldots, \widetilde{\boldsymbol{n}}^{T_1}] \in{\mathbb{C}^{T_3 \times T_1}}.
\end{split}
\end{equation}
Then we have
\begin{equation}\label{Ru}
\boldsymbol{R}_u^k = \boldsymbol{W}^k \boldsymbol{H}_u^k \boldsymbol{F}_u^k + \widetilde{\boldsymbol{n}}.
\end{equation}
We need to estimate $\boldsymbol{H}_u^k$ based on $\boldsymbol{R}_u^k$, $\boldsymbol{W}^k$, and $\boldsymbol{F}_u^k$ for all $k$~\cite{he2018geometrical}, which will be discussed in the following section.

\section{TDE-based Channel Estimation Scheme}
In this section, we propose the TDE-based channel estimation scheme by obtaining a high-resolution estimate of the AoA and AoD in three stages. In the first stage, we use all $K$ subcarriers to transmit pilot signal. In the second and third stages, we only need to use one subcarrier, i.e., the first one $(k=0)$, to transmit pilot signal while using the remaining $K-1$ subcarriers to transmit data, which is in the same fashion as the current LTE system with pilot subcarriers embedded in the data subcarriers. We use the received pilot sequences at the subcarrier $k=0$ in the first and second stages to estimate the AoA, which is addressed in detail in Section~III.A. Then we use the received pilot sequences at the subcarrier $k=0$ in the first and third stages to estimate the AoD in Section~III.B. Finally, the AoA and AoD are paired  and the channels for all $K$ subcarriers are reconstructed using the pilot sequences transmitted at all $K$ subcarriers in the first stage in Section~III.C.

\subsection{ AoA Estimation}
In the first stage, the BS and $U$ users turn off the $N_A$th and the $M_A$th antennas (the last antennas), respectively. Note that there is a great difference between the proposed channel estimation methods and that in~\cite{liao20172d} except the wideband or narrowband channels. Unlike the existing ESPRIT-based or MUSIC-based channel estimation schemes that have to turn off approximately half of the antennas so that the number of active antennas is equal to that of time slots for channel
estimation~\cite{liao20172d,guo2017millimeter}, here we turn off only one antenna at each side, which ensures almost the same of transmission power and signal coverage. Note that if each user only has $M_A=1$ antenna, i.e., the MISO scenario, $U$ users do not need to turn off the single antenna because the AoD is nonexistent in this case.

We use all $K$ subcarriers to transmit pilot sequences for $T$ time slots. To distinguish hybrid precoding and combining matrices in these three stages, we denote $\boldsymbol{W}^k$ in (\ref{DefinitionW}) and $\boldsymbol{F}_u^k$ in (\ref{DefinitionF}) with $k=0$ in the first stage as $\widetilde{\boldsymbol{W}}^{(1)}$ and $\widetilde{\boldsymbol{F}}_u^{(1)}$, respectively, which can be represented as
\begin{align}\label{Definition_WF}
  \widetilde{\boldsymbol{W}}^{(1)} &= [\widetilde{\boldsymbol{W}}, \boldsymbol{0}^{T_3}], ~~ \widetilde{\boldsymbol{F}}_u^{(1)} = [(\widetilde{\boldsymbol{F}}_u)^T,  (\boldsymbol{0}^{T_1})^T]^T,
\end{align}
where $\widetilde{\boldsymbol{W}} \in{\mathbb{C}^{T_3 \times{(N_A-1)}}}$ and $\widetilde{\boldsymbol{F}}_u \in{\mathbb{C}^{(M_A-1) \times T_1}}$ denote the hybrid combining and precoding matrices connected to the powered $N_A-1$ and $M_A-1$ antennas, respectively. Combining (\ref{Ru}), (\ref{Huk2}) and (\ref{Definition_WF}), we have
\begin{align}\label{Ru01}
\widetilde{\boldsymbol{R}}_u^{(1)} &= \widetilde{\boldsymbol{W}}^{(1)} \boldsymbol{H}_u^0 \widetilde{\boldsymbol{F}}_u^{(1)} + \widetilde{\boldsymbol{n}} = \gamma  \widetilde{\boldsymbol{W}}^{(1)} \boldsymbol{A}_{u,R} \boldsymbol{\Lambda}_u^0 \boldsymbol{A}_{u,T}^H \widetilde{\boldsymbol{F}}_u^{(1)} + \widetilde{\boldsymbol{n}} \notag \\ &= \gamma  \widetilde{\boldsymbol{W}} \boldsymbol{A}_{u,R}^1 \boldsymbol{\Lambda}_u^0 (\boldsymbol{A}_{u,T}^1)^H \widetilde{\boldsymbol{F}}_u + \widetilde{\boldsymbol{n}},
\end{align}
for $k=0$, where $\boldsymbol{A}_{u,R}^1 \in{\mathbb{C}^{(N_A-1)\times L_u}}$ and $\boldsymbol{A}_{u,T}^1 \in{\mathbb{C}^{(M_A-1) \times L_u}}$ consist of the first $N_A-1$ and $M_A-1$ rows of $\boldsymbol{A}_{u,R}$ and $\boldsymbol{A}_{u,T}$, respectively, which are denoted as
\begin{align}\label{AuR1}
\boldsymbol{A}_{u,R}^1 &= \frac{1}{\sqrt{N_A}}
    \begin{bmatrix}
        1     & \cdots & 1 \\
        e^{j\pi\theta_{u,1}}    & \cdots &  e^{j\pi\theta_{u,L_u}}      \\
        \vdots & \ddots & \vdots\\
        e^{j\pi\theta_{u,1}(N_A-2)}  &   \cdots & e^{j\pi\theta_{u,L_u}(N_A-2)}
    \end{bmatrix},  \notag \\
\boldsymbol{A}_{u,T}^1 &= \frac{1}{\sqrt{M_A}}
    \begin{bmatrix}
        1     & \cdots & 1 \\
        e^{j\pi\phi_{u,1}}   & \cdots &  e^{j\pi\phi_{u,L_u}}     \\
        \vdots & \ddots & \vdots\\
        e^{j\pi\phi_{u,1}(M_A-2)}    & \cdots & e^{j\pi\phi_{u,L_u}(M_A-2)}
    \end{bmatrix}.
\end{align}

In the second stage, the BS and $U$ users turn off the $1$st and the $M_A$th antennas, respectively. Note that we only turn off one antenna at each side. However, in this stage, we only need to use the subcarrier $k=0$ to transmit pilot sequences for $T$ time slots. Note that the remaining $K-1$ subcarriers are used to transmit data, which can be recovered immediately after the channel estimation. $\widetilde{\boldsymbol{F}}_u^{(1)}$ is set to be the same as that in (\ref{Definition_WF}). Different from the first stage, $\boldsymbol{W}^k$ in this stage is denoted as
\begin{equation}
  \widetilde{\boldsymbol{W}}^{(2)} = [\boldsymbol{0}^{T_3}, \widetilde{\boldsymbol{W}}]
\end{equation}
where $\widetilde{\boldsymbol{W}}$ is set to be the same as that in (\ref{Definition_WF}). Therefore, (\ref{Ru}) can be represented as
\begin{align}\label{Ru02}
\widetilde{\boldsymbol{R}}_u^{(2)} &= \gamma  \widetilde{\boldsymbol{W}}^{(2)} \boldsymbol{A}_{u,R} \boldsymbol{\Lambda}_u^0 \boldsymbol{A}_{u,T}^H \widetilde{\boldsymbol{F}}_u^{(1)} + \widetilde{\boldsymbol{n}} \notag \\ &= \gamma  \widetilde{\boldsymbol{W}} \boldsymbol{A}_{u,R}^2 \boldsymbol{\Lambda}_u^0 (\boldsymbol{A}_{u,T}^1)^H \widetilde{\boldsymbol{F}}_u + \widetilde{\boldsymbol{n}}
\end{align}
where $\boldsymbol{A}_{u,R}^2 \in{\mathbb{C}^{(N_A-1)\times L_u}}$ is consisted of the last $N_A-1$ rows of $\boldsymbol{A}_{u,R}$ as
\begin{align}\label{AuR2}
\boldsymbol{A}_{u,R}^2 &= \frac{1}{\sqrt{N_A}}
    \begin{bmatrix}
        e^{j\pi\theta_{u,1}}   & \cdots &  e^{j\pi\theta_{u,L_u}}      \\
        e^{j\pi\theta_{u,1}2}    & \cdots &  e^{j\pi\theta_{u,L_u}2}      \\
        \vdots & \ddots & \vdots\\
        e^{j\pi\theta_{u,1}(N_A-1)}     & \cdots & e^{j\pi\theta_{u,L_u}(N_A-1)}
    \end{bmatrix}.
\end{align}
Based on (\ref{AuR1}) and (\ref{AuR2}), we have
\begin{equation}\label{AuRTheta}
  \boldsymbol{A}_{u,R}^2 = \boldsymbol{A}_{u,R}^1 \boldsymbol{\Theta}
\end{equation}
where $\boldsymbol{\Theta} \in{\mathbb{C}^{L_u \times L_u}}$ is the diagonal matrix denoted as
\begin{equation}
  \boldsymbol{\Theta} = \textrm{diag}\left( [ e^{j\pi\theta_{u,1}},e^{j\pi\theta_{u,2}},\ldots,e^{j\pi\theta_{u,L_u}} ] \right).
\end{equation}

\begin{algorithm}[!t]
	\caption{AoA Estimation for the TDE-based Channel Estimation Scheme}
	\label{alg1}
	\begin{algorithmic}[1]
		\STATE \emph{Input:} $\widetilde{\boldsymbol{R}}_u^{(1)}$, $\widetilde{\boldsymbol{R}}_u^{(2)}$.

        \STATE Obtain $\widetilde{\boldsymbol{R}}_u$ via (\ref{barRuk}).
        \STATE Compute $\boldsymbol{B}$ via (\ref{B}).
        \STATE Obtain $\boldsymbol{U}$ via (\ref{U}).
        \STATE Compute $\boldsymbol{U}_{s,1}$ and $\boldsymbol{U}_{s,2}$ via (\ref{U1U2}).
        \STATE Obtain $\boldsymbol{\Psi}$ via (\ref{Psi}).
        \STATE Compute $L_u$ eigenvalues of $\boldsymbol{\Psi}$ as $\{\lambda_i,i=1,2,\ldots,L_u\}$.
        \STATE Obtain $\hat{\theta}_{u,i}$ via (\ref{hat_theta}), $i=1,2,\ldots,L_u$.

        \STATE \emph{Output:} $\hat{\theta}_{u,i}$, $i=1,2,\ldots,L_u$.
	\end{algorithmic}
\end{algorithm}

Identity (\ref{AuRTheta}) shows the rotation invariance property of the channel steering vectors, which is typically employed by the ESPRIT method. Our Algorithm~\ref{alg1} is based on the ESPRIT method~\cite{roy1989esprit}. As shown in Algorithm~\ref{alg1}, we use $\widetilde{\boldsymbol{R}}_u^{(1)}$ and $\widetilde{\boldsymbol{R}}_u^{(2)}$ to obtain an estimate of $\theta_{u,i}, i=1,2,\ldots,L_u$. By stacking $\widetilde{\boldsymbol{R}}_u^{(1)}$ and $\widetilde{\boldsymbol{R}}_u^{(2)}$ at step~2, we define a matrix $\widetilde{\boldsymbol{R}}_u \in{\mathbb{C}^{2T_3 \times T_1}}$ as
\begin{align}\label{barRuk}
  \widetilde{\boldsymbol{R}}_u =
  \begin{bmatrix}
    \widetilde{\boldsymbol{R}}_u^{(1)} \\
    \widetilde{\boldsymbol{R}}_u^{(2)}
  \end{bmatrix}
  =
  \boldsymbol{B}_R \boldsymbol{B}_T + \bar{\boldsymbol{n}},
\end{align}
where
\begin{align}
  \boldsymbol{B}_R &\triangleq
  \begin{bmatrix}
    \gamma  \widetilde{\boldsymbol{W}} \boldsymbol{A}_{u,R}^1  \\
    \gamma  \widetilde{\boldsymbol{W}} \boldsymbol{A}_{u,R}^1 \boldsymbol{\Theta}
  \end{bmatrix} \in{\mathbb{C}^{2T_3 \times L_u}}, \notag \\
  \boldsymbol{B}_T &\triangleq
  \boldsymbol{\Lambda}_u^0 (\boldsymbol{A}_{u,T}^1)^H \widetilde{\boldsymbol{F}}_u \in{\mathbb{C}^{L_u \times T_1}},
\end{align}
and $\bar{\boldsymbol{n}} \in{\mathbb{C}^{2T_3 \times T_1}}$ is the additive noise by stacking the noise term in $\widetilde{\boldsymbol{R}}_u^{(1)}$ and $\widetilde{\boldsymbol{R}}_u^{(2)}$. Define $\boldsymbol{b}_{T,t}$ and $\bar{\boldsymbol{n}}_t$ as the $t$th column of $\boldsymbol{B}_T$ and $\bar{\boldsymbol{n}}$, respectively. Denote
\begin{align}\label{B}
  \boldsymbol{B}  &\triangleq \widetilde{\boldsymbol{R}}_u (\widetilde{\boldsymbol{R}}_u)^H  = \sum_{t=1}^{T_1} (\boldsymbol{B}_R \boldsymbol{b}_{T,t}+\bar{\boldsymbol{n}}_t)(\boldsymbol{B}_R \boldsymbol{b}_{T,t}+\bar{\boldsymbol{n}}_t)^H \notag \\ &\approx  \boldsymbol{B}_R (\boldsymbol{B}_T \boldsymbol{B}_T^H) \boldsymbol{B}_R^H + \bar{\boldsymbol{n}}\bar{\boldsymbol{n}}^H.
\end{align}
Note that the correlation between $\boldsymbol{B}_R \boldsymbol{B}_T$ and $\bar{\boldsymbol{n}}$ is greatly weakened after $T_1$ additions. At step~4 in Algorithm~\ref{alg1}, the singular value decomposition (SVD) of the positive semi-definite $\boldsymbol{B}$ can be represented as
\begin{equation}\label{U}
  \boldsymbol{B}=\boldsymbol{U}\boldsymbol{\Sigma} \boldsymbol{U}^H,
\end{equation}
where $\boldsymbol{U}\in{\mathbb{C}^{2T_3\times{2T_3}}}$ is a unitary matrix and $\boldsymbol{\Sigma}\in{\mathbb{C}^{2T_3\times{2T_3}}}$ is a real diagonal matrix with diagonal entries sorted in
descending order. Since there are $L_u$ paths, the first $L_u$ columns of $\boldsymbol{U}$ corresponding to the $L_u$ largest diagonal entries of $\boldsymbol{\Sigma}$ form the signal subspace $\boldsymbol{U}_s$. Since $\boldsymbol{B}_R$ and $\boldsymbol{U}_s$ share the same basis of their respective $L_u$ column vectors, there exists an invertible matrix $\boldsymbol{T}_R \in{\mathbb{C}^{L_u\times{L_u}}}$ satisfying $\boldsymbol{U}_s=\boldsymbol{B}_R \boldsymbol{T}_R$. Dividing $\boldsymbol{U}_s$ into two submatrices $\boldsymbol{U}_{s,1} \in{\mathbb{C}^{T_3 \times L_u}}$ and $\boldsymbol{U}_{s,2} \in{\mathbb{C}^{T_3 \times L_u}}$ at step~5, we have
\begin{align}\label{U1U2}
  \boldsymbol{U}_s =
  \begin{bmatrix}
    \boldsymbol{U}_{s,1} \\
    \boldsymbol{U}_{s,2}
  \end{bmatrix}
  =
  \begin{bmatrix}
    \gamma  \widetilde{\boldsymbol{W}} \boldsymbol{A}_{u,R}^1  \\
    \gamma  \widetilde{\boldsymbol{W}} \boldsymbol{A}_{u,R}^1 \boldsymbol{\Theta}
  \end{bmatrix} \boldsymbol{T}_R
  =
  \begin{bmatrix}
    \gamma  \widetilde{\boldsymbol{W}} \boldsymbol{A}_{u,R}^1 \boldsymbol{T}_R \\
    \gamma  \widetilde{\boldsymbol{W}} \boldsymbol{A}_{u,R}^1 \boldsymbol{\Theta} \boldsymbol{T}_R
  \end{bmatrix}.
\end{align}
Since $\boldsymbol{T}_R$ is an invertible matrix, we have
\begin{equation}
  \boldsymbol{U}_{s,2} = \boldsymbol{U}_{s,1} \boldsymbol{T}_R^{-1} \boldsymbol{\Theta} \boldsymbol{T}_R.
\end{equation}
Define $\boldsymbol{\Psi} \triangleq \boldsymbol{T}_R^{-1} \boldsymbol{\Theta} \boldsymbol{T}_R \in{\mathbb{C}^{L_u \times L_u}}$, i.e., the eigenvalue decomposition of $\boldsymbol{\Psi}$, where $\boldsymbol{T}_R$ consists of $L_u$ eigenvectors and $\boldsymbol{\Theta}$ consists of $L_u$ eigenvalues on the diagonal. We obtain $\boldsymbol{\Psi}$ at step~6 as
\begin{equation}\label{Psi}
  \boldsymbol{\Psi} = \left(\boldsymbol{U}_{s,1}\right)^{\dag} \boldsymbol{U}_{s,2} = \left((\boldsymbol{U}_{s,1})^H\boldsymbol{U}_{s,1}\right)^{-1}(\boldsymbol{U}_{s,1})^H \boldsymbol{U}_{s,2}.
\end{equation}
To guarantee the existence of $(\boldsymbol{U}_{s,1})^{\dag}$, we only require $T_3 \ge L_u$. Define the $L_u$ eigenvalues of $\boldsymbol{\Psi}$ as $\{\lambda_i,i=1,2,\ldots,L_u\}$. Then the estimated AoA of the $i$th path for the $u$th user can be expressed at step~8 as
\begin{equation}\label{hat_theta}
  \hat{\theta}_{u,i} = \frac{\textrm{arg} ( \lambda_i )}{\pi}
\end{equation}
for $i = 1,2,\ldots,L_u$, where $\textrm{arg}(\lambda)\in [-\pi,\pi)$ denotes the phase angle of the complex number $\lambda$. Finally we output the estimated AoA of the $i$th path for the $u$th user at step~9.

\subsection{ AoD Estimation}
In the third stage, the BS and $U$ users turn off the $N_A$th and the $1$st antennas, respectively. Similarly, in this stage, we only need to use the subcarrier $k=0$ to transmit pilot sequences for $T$ time slots and use the remaining $K-1$ subcarriers to transmit data. $\widetilde{\boldsymbol{W}}^{(1)}$ is set to be the same as that in (\ref{Definition_WF}).  Different from the first stage, $\boldsymbol{F}_u^k$ in the third stage is denoted as
\begin{equation}
  \widetilde{\boldsymbol{F}}_u^{(2)} = [(\boldsymbol{0}^{T_1})^T, (\widetilde{\boldsymbol{F}}_u)^T]^T,
\end{equation}
where $\widetilde{\boldsymbol{F}}_u$ is set to be the same as that in (\ref{Definition_WF}). Therefore, (\ref{Ru}) can be represented as
\begin{align}
\widetilde{\boldsymbol{R}}_u^{(3)} &= \gamma  \widetilde{\boldsymbol{W}}^{(1)} \boldsymbol{A}_{u,R} \boldsymbol{\Lambda}_u^0 \boldsymbol{A}_{u,T}^H \widetilde{\boldsymbol{F}}_u^{(2)} + \widetilde{\boldsymbol{n}} \notag \\ &= \gamma  \widetilde{\boldsymbol{W}} \boldsymbol{A}_{u,R}^1 \boldsymbol{\Lambda}_u^0 (\boldsymbol{A}_{u,T}^2)^H \widetilde{\boldsymbol{F}}_u + \widetilde{\boldsymbol{n}},
\end{align}
where $\boldsymbol{A}_{u,T}^2 \in{\mathbb{C}^{(M_A-1)\times L_u}}$ consists of the last $M_A-1$ rows of $\boldsymbol{A}_{u,T}$ as
\begin{align}\label{AuT2}
\boldsymbol{A}_{u,T}^2 &= \frac{1}{\sqrt{M_A}}
    \begin{bmatrix}
        e^{j\pi\phi_{u,1}}     & \cdots &  e^{j\pi\phi_{u,L_u}}      \\
        e^{j\pi\phi_{u,1}2}     & \cdots &  e^{j\pi\phi_{u,L_u}2}      \\
        \vdots & \ddots & \vdots\\
        e^{j\pi\phi_{u,1}(M_A-1)}     & \cdots & e^{j\pi\phi_{u,L_u}(M_A-1)}
    \end{bmatrix}.
\end{align}
Based on (\ref{AuR1}) and (\ref{AuT2}), we have
\begin{equation}\label{AuTPhi}
  \boldsymbol{A}_{u,T}^2 = \boldsymbol{A}_{u,T}^1 \boldsymbol{\Phi},
\end{equation}
where $\boldsymbol{\Phi} \in{\mathbb{C}^{L_u \times L_u}}$ is the diagonal matrix denoted as
\begin{equation}
  \boldsymbol{\Phi} = \textrm{diag}\left( [ e^{j\pi\phi_{u,1}},e^{j\pi\phi_{u,2}},\ldots,e^{j\pi\phi_{u,L_u}} ] \right).
\end{equation}
Since (\ref{AuTPhi}) has the same structure as (\ref{AuRTheta}), we can directly use Algorithm~\ref{alg1} to obtain the estimated AoD of the $i$th path of the $u$th user $\hat{\phi}_{u,i}$ by replacing
\begin{align}\notag
  \big[   \widetilde{\boldsymbol{R}}_u^{(1)}, \widetilde{\boldsymbol{R}}_u^{(2)}, \widetilde{\boldsymbol{W}}, \boldsymbol{A}_{u,R}^1, \boldsymbol{\Theta}, \boldsymbol{\Lambda}_u^0, (\boldsymbol{A}_{u,T}^1)^H, \widetilde{\boldsymbol{F}}_u, T_3, T_1, \hat{\theta}_{u,i}  \big] \notag
\end{align}
with
\begin{align}\notag
  \big[   &(\widetilde{\boldsymbol{R}}_u^{(1)})^H, (\widetilde{\boldsymbol{R}}_u^{(3)})^H, (\widetilde{\boldsymbol{F}}_u)^H, \boldsymbol{A}_{u,T}^1, \boldsymbol{\Phi}, \notag \\  &(\boldsymbol{\Lambda}_u^0)^H, (\boldsymbol{A}_{u,R}^1)^H, (\widetilde{\boldsymbol{W}})^H, T_1, T_3, \hat{\phi}_{u,i}  \big]. \notag
\end{align}

\subsection{ AoA and AoD Pairing}
Once $\theta_{u,i}$ and $\phi_{u,i}$ are estimated, we can obtain the estimation of $\boldsymbol{A}_{u,R}$ and $\boldsymbol{A}_{u,T}$ as
\begin{align}\label{hat_AuRT}
\hat{\boldsymbol{A}}_{u,R} &= [\boldsymbol{\alpha} (N_A,\hat{\theta}_{u,1}),\boldsymbol{\alpha} (N_A,\hat{\theta}_{u,2}),\ldots,\boldsymbol{\alpha} (N_A,\hat{\theta}_{u,L_u})],  \notag \\
\hat{\boldsymbol{A}}_{u,T} &= [\boldsymbol{\alpha} (M_A,\hat{\phi}_{u,1}),\boldsymbol{\alpha} (M_A,\hat{\phi}_{u,2}),\ldots,\boldsymbol{\alpha} (M_A,\hat{\phi}_{u,L_u})].
\end{align}
Since each channel path corresponds to an AoA and an AoD that are in pairs, incorrectly pairing of the channel AoAs and AoDs will lead to a completely different channel.

Define $\boldsymbol{v}_u^k \triangleq [\boldsymbol{\Lambda}_u^k[1,1],\boldsymbol{\Lambda}_u^k[2,2],\ldots,\boldsymbol{\Lambda}_u^k[L_u,L_u]]^T \in{\mathbb{C}^{L_u}}$. Then $\boldsymbol{R}_u^k$ in (\ref{Ru}) obtained in the first stage can be represented in vector form as
\begin{align}\label{vecRuk}
\textrm{vec}(\boldsymbol{R}_u^k) &= \textrm{vec}(\boldsymbol{W}^k \boldsymbol{H}_u^k \boldsymbol{F}_u^k) + \textrm{vec}(\widetilde{\boldsymbol{n}})  \\ &\overset{(a)}= \left((\boldsymbol{F}_u^k)^T \otimes \boldsymbol{W}^k \right)\textrm{vec}(\boldsymbol{H}_u^k)  + \textrm{vec}(\widetilde{\boldsymbol{n}}) \notag \\
&\overset{(b)}= \gamma\left((\boldsymbol{F}_u^k)^T \otimes \boldsymbol{W}^k \right)\left(\boldsymbol{A}_{u,T}^* \circ \boldsymbol{A}_{u,R} \right)\boldsymbol{v}_u^k  + \textrm{vec}(\widetilde{\boldsymbol{n}})\notag
\end{align}
where $(\cdot)^* $ denotes the conjugate, the equality marked by (a) holds due to the fact that $\textrm{vec}(\boldsymbol{ABC}) = (\boldsymbol{C}^T \otimes \boldsymbol{A}) \textrm{vec}(\boldsymbol{B})$, and the equality marked by (b) follows from the channel model in (\ref{Huk2}) and the properties of the Khatri-Rao product. Then $\boldsymbol{v}_u^k$ can be obtained via LS estimation as
\begin{equation}\label{hatbardelta}
  \hat{\boldsymbol{v}}_u^k = \frac{1}{\gamma} \left( \left((\boldsymbol{F}_u^k)^T \otimes \boldsymbol{W}^k \right)\left(\hat{\boldsymbol{A}}_{u,T}^* \circ \hat{\boldsymbol{A}}_{u,R} \right) \right)^\dag \textrm{vec}(\boldsymbol{R}_u^k).
\end{equation}
However, it requires that the $L_u$ columns of $\hat{\boldsymbol{A}}_{u,R}$ and $\hat{\boldsymbol{A}}_{u,T}$ are paired, i.e., $\hat{\theta}_{u,i}$ and $\hat{\phi}_{u,i}$ are paired for $i=1,2,\ldots,L_u$. Otherwise $\left(\hat{\boldsymbol{A}}_{u,T}^* \circ \hat{\boldsymbol{A}}_{u,R} \right)  \ne  \left(\boldsymbol{A}_{u,T}^* \circ \boldsymbol{A}_{u,R} \right)$, resulting in large estimation error of $\hat{\boldsymbol{v}}_u^k$. Therefore we should pair $\hat{\theta}_{u,i}$ and $\hat{\phi}_{u,i}$ before estimating $\boldsymbol{v}_u^k$.

Since all $K$ subcarriers share the same AoA and AoD, we can estimate the AoA and AoD based on the first subcarrier, i.e., $k=0$. Then $\boldsymbol{\Lambda}_u^k$ in \eqref{Huk2} on the first subcarrier can be estimated as
\begin{equation}\label{hat_bar_Delta}
  \hat{\boldsymbol{\Lambda}}_u^0 = \frac{1}{\gamma} (\widetilde{\boldsymbol{W}}^{(1)} \hat{\boldsymbol{A}}_{u,R})^{\dag} \widetilde{\boldsymbol{R}}_u^{(1)}     (\hat{\boldsymbol{A}}_{u,T}^H \widetilde{\boldsymbol{F}}_u^{(1)})^{\dag}.
\end{equation}
Note that $\boldsymbol{\Lambda}_u^0$ is a diagonal matrix with $L_u$ nonzero entries according to \eqref{Huk2}. However, if $\hat{\theta}_{u,i}$ and $\hat{\phi}_{u,i}$ are incorrectly paired, $\hat{\boldsymbol{\Lambda}}_u^0$ will not be diagonal. Therefore, the structure of $\hat{\boldsymbol{\Lambda}}_u^0$ needs to be analyzed.

Denote
\begin{align}
  \boldsymbol{\theta}_{u} &\triangleq [\theta_{u,1},\theta_{u,2},\ldots,\theta_{u,L_u}],
  ~~\boldsymbol{\phi}_{u} \triangleq [\phi_{u,1},\phi_{u,2},\ldots,\phi_{u,L_u}], \notag \\
  \hat{\boldsymbol{\theta}}_{u} &\triangleq [\hat{\theta}_{u,1},\hat{\theta}_{u,2},\ldots,\hat{\theta}_{u,L_u}],
  ~~\hat{\boldsymbol{\phi}}_{u} \triangleq [\hat{\phi}_{u,1},\hat{\phi}_{u,2},\ldots,\hat{\phi}_{u,L_u}].
\end{align}
It is obvious that $\hat{\boldsymbol{\theta}}_{u}$ and $\hat{\boldsymbol{\phi}}_{u}$ are the permutations of $\boldsymbol{\theta}_{u}$ and $\boldsymbol{\phi}_{u}$, respectively, if neglecting the additive noise. Define $\bar{\boldsymbol{p}} \in{\mathbb{Z}^{L_u}}$ and $\boldsymbol{q} \in{\mathbb{Z}^{L_u}}$ as two permutations of $\{1,2,\ldots,L_u\}$. Then we have
\begin{align}\label{theta_q}
 \hat{\theta}_{u,i} = \theta_{u,\bar{\boldsymbol{p}}[i]}, ~~\hat{\phi}_{u,i} = \phi_{u,\boldsymbol{q}[i]}.
\end{align}
Denote $\boldsymbol{C}_{\bar{\boldsymbol{p}}}$ to be a $L_u \times L_u$ square matrix where only $\boldsymbol{C}_{\bar{\boldsymbol{p}}} \big[\bar{\boldsymbol{p}}[i],i\big]=1, i=1,2,\ldots,L_u$ and all the other entries are zero. Then (\ref{theta_q}) can be expressed in matrix form as
\begin{align}
 \hat{\boldsymbol{\theta}}_{u} = \boldsymbol{\theta}_{u} \boldsymbol{C}_{\bar{\boldsymbol{p}}}, ~~\hat{\boldsymbol{\phi}}_{u} = \boldsymbol{\phi}_{u}\boldsymbol{C}_{\boldsymbol{q}}.
\end{align}
Similarly, (\ref{hat_AuRT}) can be further expressed as
\begin{align}\label{AuR3}
\hat{\boldsymbol{A}}_{u,R} = \boldsymbol{A}_{u,R} \boldsymbol{C}_{\bar{\boldsymbol{p}}}, ~~\hat{\boldsymbol{A}}_{u,T} = \boldsymbol{A}_{u,T} \boldsymbol{C}_{\boldsymbol{q}}.
\end{align}
Denote $\boldsymbol{Z}_{p,q}$ to be a $L_u \times L_u$ square matrix where only $\boldsymbol{Z}_{p,q}[p,q]=1$ and all the other entries are zero. Combining (\ref{hat_bar_Delta}) and (\ref{AuR3}), we have
\begin{align}
  \hat{\boldsymbol{\Lambda}}_u^0 &= (\boldsymbol{C}_{\bar{\boldsymbol{p}}})^{-1} \boldsymbol{\Lambda}_u^0 ((\boldsymbol{C}_{\boldsymbol{q}})^H)^{-1}  \overset{(a)} = (\boldsymbol{C}_{\bar{\boldsymbol{p}}})^H \boldsymbol{\Lambda}_u^0 \boldsymbol{C}_{\boldsymbol{q}}  \notag \\ & = \sum_{i=1}^{L_u} \boldsymbol{\Lambda}_u^0[i,i] \boldsymbol{Z}_{\boldsymbol{p}[i],\boldsymbol{q}[i]}
\end{align}
where the equality marked by (a) holds due to the fact that $(\boldsymbol{C}_{\bar{\boldsymbol{p}}})^{-1} = (\boldsymbol{C}_{\bar{\boldsymbol{p}}})^H$ and $\boldsymbol{p}\in{\mathbb{Z}^{L_u}}$ is a permutation of $\{1,2,\ldots,L_u\}$. $\hat{\boldsymbol{\Lambda}}_u^0$ is a permutation of $\boldsymbol{\Lambda}_u^0$,  where $(\boldsymbol{p}[i],\boldsymbol{q}[i])$ represents the coordinate of $\boldsymbol{\Lambda}_u^0[i,i]$. Furthermore, there is only one nonzero entry in each row and each column of $\hat{\boldsymbol{\Lambda}}_u^0$. By searching the coordinates of these $L_u$ nonzero entries, we can pair $\boldsymbol{p}$ and $\boldsymbol{q}$ with the same indices, i.e., $(\boldsymbol{p}[i], \boldsymbol{q}[i]), i=1,2,\ldots,L_u$. Define $\hat{\boldsymbol{p}}$ and $\hat{\boldsymbol{q}}$ as the estimate of $\boldsymbol{p}$ and $\boldsymbol{q}$, respectively. Then we can obtain the paired AoA and AoD of the $L_u$ paths for the $u$th user as
\begin{equation}\label{pair_theta}
 \tilde{\boldsymbol{\theta}}_{u} = \hat{\boldsymbol{\theta}}_{u}(\boldsymbol{C}_{\bar{\boldsymbol{p}}})^{-1} = \hat{\boldsymbol{\theta}}_{u}\boldsymbol{C}_{\hat{\boldsymbol{p}}}, ~~\tilde{\boldsymbol{\phi}}_{u} = \hat{\boldsymbol{\phi}}_{u}(\boldsymbol{C}_{\hat{\boldsymbol{q}}})^{-1}.
\end{equation}

\begin{algorithm}[!t]
	\caption{AoA and AoD Pairing for the TDE-based Channel Estimation Scheme}
	\label{alg2}
	\begin{algorithmic}[1]
		\STATE \emph{Input:} $\hat{\boldsymbol{\theta}}_{u}$, $\hat{\boldsymbol{\phi}}_{u}$, $\hat{\boldsymbol{\Lambda}}_u^0$.
        \STATE Initialization: $\hat{\boldsymbol{p}} \leftarrow \boldsymbol{0}^{L_u}$. $\hat{\boldsymbol{q}} \leftarrow \boldsymbol{0}^{L_u}$. $\boldsymbol{\Gamma}_1 \leftarrow \{1,2,\ldots,L_u\}$.

        \FOR{$i=1,2,\ldots,L_u$}
            \STATE Obtain $q_i$ via (\ref{qi}).
            \STATE Obtain the $i$th entry of $\hat{\boldsymbol{p}}$ and $\hat{\boldsymbol{q}}$ via (\ref{hat_pi}).
            \STATE Update $\boldsymbol{\Gamma}_{i+1}$ via (\ref{Gamma1}).
        \ENDFOR
        \STATE Obtain $\tilde{\boldsymbol{\theta}}_{u}$ and $\tilde{\boldsymbol{\phi}}_{u}$ via (\ref{pair_theta}).

        \STATE \emph{Output:} $\tilde{\boldsymbol{\theta}}_{u}$, $\tilde{\boldsymbol{\phi}}_{u}$.
	\end{algorithmic}
\end{algorithm}

Based on the analysis of the structure of $\hat{\boldsymbol{\Lambda}}_u^0$, now we propose Algorithm~\ref{alg2} to pair the AoA and AoD of the $L_u$ paths for the $u$th user. Define $\boldsymbol{\Gamma}_i$ as a set of column indices with $L_u-i+1$ entries, where $\boldsymbol{\Gamma}_1$ is initialized to be $\{1,2,\ldots,L_u\}$. For the $i$th iteration, we obtain the entry with the largest absolute value from the $i$th row and $d$th column of $\hat{\boldsymbol{\Lambda}}_u^0$ as
\begin{equation}\label{qi}
  q_i = \arg\underset{d\in\boldsymbol{\Gamma}_i} {\max} \big| \hat{\boldsymbol{\Lambda}}_u^0[i,d] \big|.
\end{equation}
Then we add the paired $i$ and $q_i$ to $\hat{\boldsymbol{p}}$ and $\hat{\boldsymbol{q}}$ respectively as
\begin{equation}\label{hat_pi}
  \hat{\boldsymbol{p}}[i] = i, ~~\hat{\boldsymbol{q}}[i] = q_i.
\end{equation}
Since there is only one nonzero entry in each column of $\hat{\boldsymbol{\Lambda}}_u^0$, we delete $q_i$ from $\boldsymbol{\Gamma}_i$ to obtain $\boldsymbol{\Gamma}_{i+1}$ for the next iteration as
\begin{equation}\label{Gamma1}
  \boldsymbol{\Gamma}_{i+1} = \boldsymbol{\Gamma}_{i} \setminus q_i.
\end{equation}
We repeat the above procedures for $L_u$ times to obtain the paired $\boldsymbol{p}$ and $\boldsymbol{q}$. From (\ref{pair_theta}), we obtain the paired AoA and AoD of the $L_u$ paths for the $u$th user. The detailed steps are summarized in Algorithm~\ref{alg2}.

With paired AoA and AoD, we can adjust the corresponding order of the columns of $\hat{\boldsymbol{A}}_{u,R}$ and $\hat{\boldsymbol{A}}_{u,T}$ to obtain $\tilde{\boldsymbol{A}}_{u,R}$ and $\tilde{\boldsymbol{A}}_{u,T}$, which is essentially to replace $\hat{\theta}_{u,i}$, $\hat{\phi}_{u,i}$ with $\tilde{\theta}_{u,i}$, $\tilde{\phi}_{u,i}$, respectively, in (\ref{hat_AuRT}). The estimation of $\boldsymbol{v}_u^k$ can be performed by replacing $\hat{\boldsymbol{A}}_{u,R}$ and $\hat{\boldsymbol{A}}_{u,T}$ with $\tilde{\boldsymbol{A}}_{u,R}$ and $\tilde{\boldsymbol{A}}_{u,T}$, respectively, in (\ref{hatbardelta}). Finally the estimated channel matrix at the $k$th subcarrier between the BS and the $ut$h user can be expressed as
\begin{equation}\label{Huk_hat}
\hat{\boldsymbol{H}}_{u}^k =  \gamma \tilde{\boldsymbol{A}}_{u,R} \textrm{diag}\left(\hat{\boldsymbol{v}}_u^k\right) \tilde{\boldsymbol{A}}_{u,T}^H.
\end{equation}

As a summary, we conclude the main procedures of the proposed TDE-based channel estimation scheme as follows. After the three stages of pilot transmission, we perform Algorithm~\ref{alg1} twice to obtain the AoA and AoD, respectively. Then we perform Algorithm~\ref{alg2} to pair the obtained AoA and AoD. We estimate the gain of the channel multipaths in \eqref{hatbardelta} and eventually finish the channel estimation by \eqref{Huk_hat}. Since the proposed TDE-based channel estimation scheme is based on the ESPRIT method and the ESPRIT method obtains the closed-form solution of AoA and AoD, high resolution of channel estimation can be achieved.

\section{EMS-based Channel Estimation Scheme}
In the TDE-based channel estimation scheme, there are three stages for pilot transmission. To further reduce the number of stages, we propose an EMS-based high-resolution channel estimation scheme, which only includes the first two stages of the TDE-based scheme. The critical difference between the EMS-based and the TDE-based schemes is the AoD estimation. Therefore, we only focus on the AoD subsequently.

\subsection{ AoD Estimation for the EMS-based Scheme}
In the first stage, we turn off the $N_A$th antenna at the BS while powering on all the antennas at each user.  Note that we turn off only one antenna at the BS, which ensures almost the same of transmission power and signal coverage. We use all $K$ subcarriers to transmit pilot sequences for $T$ time slots. Therefore (\ref{Ru01}) can be rewritten as
\begin{align}\label{tildeRuOne}
\widetilde{\boldsymbol{R}}_u^{(1)} &= \widetilde{\boldsymbol{W}}^{(1)} \boldsymbol{H}_u^0 \boldsymbol{F}_u^0 + \widetilde{\boldsymbol{n}} = \gamma  \widetilde{\boldsymbol{W}}^{(1)} \boldsymbol{A}_{u,R} \boldsymbol{\Lambda}_u^0 \boldsymbol{A}_{u,T}^H \boldsymbol{F}_u^0 + \widetilde{\boldsymbol{n}} \notag \\ &= \gamma  \widetilde{\boldsymbol{W}} \boldsymbol{A}_{u,R}^1 \boldsymbol{\Lambda}_u^0 \boldsymbol{A}_{u,T}^H \boldsymbol{F}_u^0 + \widetilde{\boldsymbol{n}}.
\end{align}

In the second stage, we turn off the $1$st antenna at BS while still powering on all the antennas at each user. In this stage, we only need to use the subcarrier $k=0$ to transmit pilot sequences for $T$ time slots. Therefore (\ref{Ru02}) can be rewritten as
\begin{align}
\widetilde{\boldsymbol{R}}_u^{(2)} &= \gamma  \widetilde{\boldsymbol{W}}^{(2)} \boldsymbol{A}_{u,R} \boldsymbol{\Lambda}_u^0 \boldsymbol{A}_{u,T}^H \boldsymbol{F}_u^0 + \widetilde{\boldsymbol{n}} \notag \\ &= \gamma  \widetilde{\boldsymbol{W}} \boldsymbol{A}_{u,R}^2 \boldsymbol{\Lambda}_u^0 \boldsymbol{A}_{u,T}^H \boldsymbol{F}_u^0 + \widetilde{\boldsymbol{n}}.
\end{align}

We can directly run Algorithm~\ref{alg1} to obtain the estimated AoA of the $i$th path for the $u$th user as $\hat{\theta}_{u,i}$, $i=1,2,\ldots,L_u$, by replacing $(\boldsymbol{A}_{u,T}^1)^H \widetilde{\boldsymbol{F}}_u$ with $\boldsymbol{A}_{u,T}^H \boldsymbol{F}_u^0$. Then we can obtain the estimation of $\boldsymbol{A}_{u,R}^1$ from (\ref{AuR1}) by replacing $\theta_{u,i}$ with $\hat{\theta}_{u,i}$. In this way, we finish the estimation of the AoA.

Although we can estimate the AoA similar to that in the TDE-based scheme, the AoD estimation is completely different, as there are only two stages in the EMS-based scheme. Now we propose Algorithm~\ref{alg3} to obtain an estimate of the AoD. To ease the notation of \eqref{tildeRuOne}, we define $\boldsymbol{D}_T \triangleq \boldsymbol{\Lambda}_u^0 \boldsymbol{A}_{u,T}^H \boldsymbol{F}_u^0 \in{\mathbb{C}^{L_u \times T_1}}$. Then an LS estimate of $\boldsymbol{D}_T$ based on \eqref{tildeRuOne} is obtained as
\begin{equation}\label{53}
  \hat{\boldsymbol{D}}_T = \frac{1}{\gamma} (\widetilde{\boldsymbol{W}} \hat{\boldsymbol{A}}_{u,R}^1)^{\dag} \widetilde{\boldsymbol{R}}_u^{(1)}.
\end{equation}
To guarantee the existence of $(\widetilde{\boldsymbol{W}} \hat{\boldsymbol{A}}_{u,R}^1)^{\dag}$, we only require $T_3 \ge L_u$.

We denote $\boldsymbol{d}_i \triangleq (\boldsymbol{\Lambda}_u^0[i,i])^* (\boldsymbol{F}_u^0)^H \boldsymbol{\alpha} (M_A,\phi_{u,i}) \in{\mathbb{C}^{ T_1}}$ as the $i$th column of $\boldsymbol{D}_T^H$. Then the estimation of the AoD can be converted into individual estimation of each channel path in terms of $\boldsymbol{\alpha} (M_A,\phi_{u,i})$, which will be addressed as follows.

Denote $\hat{\boldsymbol{d}}_i $ as the $i$th column of $\hat{\boldsymbol{D}}_T^H$. Denote $\boldsymbol{U}_{i}\in{\mathbb{C}^{T_1\times{(T_1-1)}}}$ to be a matrix with all $T_1-1$ pairwise orthogonal columns and orthogonal to $\hat{\boldsymbol{d}}_i $, e.g., $\boldsymbol{U}_{i}$ can be obtained by finding the null space of $\hat{\boldsymbol{d}}_i $ via SVD or Schimidt orthogonalization. We define
\begin{equation}\label{Pphi}
  \mathcal{P}_i(\phi) \triangleq  (\boldsymbol{d}(\phi))^H \boldsymbol{U}_{i} (\boldsymbol{U}_{i})^H \boldsymbol{d}(\phi),
\end{equation}
where $\boldsymbol{d}(\phi) \triangleq (\boldsymbol{F}_u^0)^H \boldsymbol{\alpha} (M_A,\phi)$. Our goal is to find an appropriate $\phi$ that minimizes $\mathcal{P}_i(\phi)$. Note that $\mathcal{P}_i(\phi)\geq 0$.

\begin{figure}[!t]
\centering
\includegraphics[width=95mm]{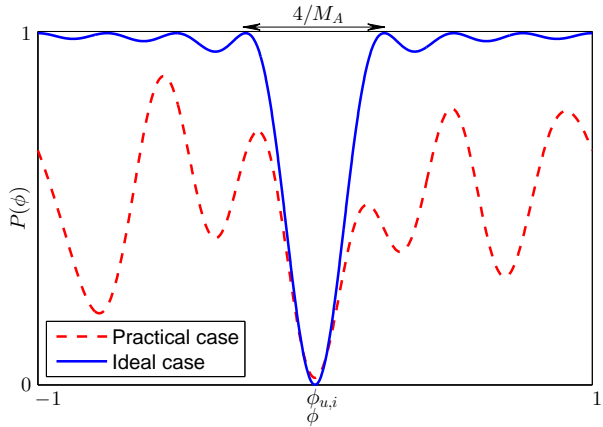}
\caption{Illustration of $\mathcal{P}_i(\phi)$ in ideal and practical cases.}
\label{Fig4}
\end{figure}

We first analyze the structure of $\mathcal{P}_i(\phi)$. To ease the analysis, we consider the ideal case without any noise and set $\boldsymbol{F}_u^0$ as a unitary matrix for simplicity. Then we have $\boldsymbol{d}_i = \hat{\boldsymbol{d}}_i = (\boldsymbol{\Lambda}_u^0[i,i])^* (\boldsymbol{F}_u^0)^H \boldsymbol{\alpha} (M_A,\phi_{u,i})$. Based on (\ref{alpha1}), (\ref{53}), and (\ref{Pphi}), direct calculation yields that
\begin{align}\label{P_phi_function_analysis}
  \mathcal{P}_i(\phi)  = 1-  \frac{ \sin^2( \pi M_A (\phi-\phi_{u,i}) /2 ) }{ {M_A}^2\sin^2( \pi (\phi-\phi_{u,i}) /2 ) }.
\end{align}
Note that in practical case with channel noise, where $\boldsymbol{d}_i \neq \hat{\boldsymbol{d}}_i$, it is difficult to analyze $\mathcal{P}_i(\phi)$. We illustrate $\mathcal{P}_i(\phi)$ for ideal and practical cases in Fig.~\ref{Fig4}, where we set $T_1=M_A/2$ in the practical case. It is seen that the practical case has the similar property as the ideal case, e.g., achieving the similar minimum by $\phi_{u,i}$ with the similar width of mainlobe of $4/M_A$.

From \eqref{P_phi_function_analysis}, $\mathcal{P}_i(\phi)$ is not monotonic and there are many local minimum points aside of the global one. Therefore, it brings challenge to the heuristic search that typically has much faster search speed than the exhaustive search. From \eqref{P_phi_function_analysis} and Fig.~\ref{Fig4}, the mainlobe of $\mathcal{P}_i(\phi)$  is $4/M_A$. If we search the minimum point within the mainlobe of $\mathcal{P}_i(\phi)$, we can find the global minimum and will not be stuck in any local minimum. Therefore, we first identify the mainlobe containing the global minimum point.

Now we propose a fast search method that first identifies the mainlobe of $\mathcal{P}_i(\phi)$ and then further searches the minimum within the mainlobe. We first identify the mainlobe of $\mathcal{P}_i(\phi)$. Since the width of the mainlobe is $4/M_A$, we sample $\mathcal{P}_i(\phi)$ with equal interval of $2/M_A$ as
\begin{equation}\label{ns}
  n_s = \arg\underset{n=1,2,\ldots,M_A} {\min}  \mathcal{P}(-1+2(n-1)/M_A).
\end{equation}
In this way, we narrow down the search space of $\phi_{u,i}$ from $[-1,1)$ to $\boldsymbol{\Gamma}$ with the width of $4/M_A$, where
\begin{align}\label{Gamma}
\boldsymbol{\Gamma} = [\Gamma_1,~&\Gamma_2], \\
\Gamma_1 \triangleq  -1+2(n_s-2)/M_A,&~~\Gamma_2 \triangleq  -1+2n_s/M_A.
\end{align}

Then we search the minimum of $\mathcal{P}_i(\phi)$ within $\boldsymbol{\Gamma}$ based on bisection search. These steps to estimate the AoD are summarized in Algorithm~\ref{alg3}. Based on the estimated AoD for the EMS-based scheme, the overall channel can be obtained similarly to the TDE-based scheme.

\begin{algorithm}[!t]
	\caption{AoD Estimation for the EMS-based Channel Estimation Scheme}
	\label{alg3}
	\begin{algorithmic}[1]
		\STATE \emph{Input:} $\widetilde{\boldsymbol{R}}_u^{(1)}$, $\widetilde{\boldsymbol{W}}$, $\hat{\boldsymbol{A}}_{u,R}^1$.
        \STATE Obtain $\hat{\boldsymbol{D}}_T$ and $\hat{\boldsymbol{d}}_i$ via (\ref{53}).
        \STATE Compute  $\mathcal{P}_i(\phi)$ via  (\ref{Pphi}).
        \STATE Obtain $n_s$ via (\ref{ns}).
        \STATE Compute $\boldsymbol{\Gamma} = [\Gamma_1,\Gamma_2]$ via (\ref{Gamma}).
        \STATE Obtain $\hat{\phi}_{u,i}$ within $\boldsymbol{\Gamma}$ via bisection search.
        \STATE \emph{Output:} $\hat{\phi}_{u,i}$.
	\end{algorithmic}
\end{algorithm}

\subsection{Computational Complexity Comparison}
Now we compare the proposed two schemes together with the SWOMP-based~\cite{javier2018frequency}, OMP-based~\cite{venugopal2017time} and DGMP-based~\cite{dai2016estimation} channel estimation schemes in terms of computational complexity.

The computational complexity of the proposed channel estimation schemes mainly comes from the channel estimation for the AoA and AoD as well as the AoA and AoD pairing in Algorithm~\ref{alg1}, Algorithm~\ref{alg2}, and Algorithm~\ref{alg3}.

For the TDE-based scheme, to estimate the AoA using Algorithm~\ref{alg1}, we need to compute $\boldsymbol{B}$ by multiplying $\widetilde{\boldsymbol{R}}_u$ and $(\widetilde{\boldsymbol{R}}_u)^H$ with their dimension of $2T_3 \times T_1$ and $T_1 \times 2T_3$ respectively, leading to the computational complexity to be $\mathcal{O}(8T_1T_3^2)$. Then we obtain $\boldsymbol{U}$ via the SVD of $\boldsymbol{B}$,  resulting in the complexity to be $\mathcal{O}(8T_3^3)$. After that, we calculate the pseudo inverse of $\boldsymbol{U}_{s,1}$ and the multiplication of $(\boldsymbol{U}_{s,1})^{\dag}$ and $\boldsymbol{U}_{s,2}$, leading to the complexity to be $\mathcal{O}(6L_u^2T_3+4L_u^3)$. Then we compute the eigenvalue of $\boldsymbol{\Psi}$ with the complexity to be $\mathcal{O}(2L_u^3)$. Therefore the total computational complexity for the estimate of the AoA is $\mathcal{O}(8T_1T_3^2+8T_3^3+6L_u^2T_3+6L_u^3)$. Similarly, the total computational complexity for the estimate of the AoD is $\mathcal{O}(8T_3T_1^2+8T_1^3+6L_u^2T_1+6L_u^3)$. To pair the AoA and AoD using Algorithm~\ref{alg2}, we need to calculate $\hat{\boldsymbol{\Lambda}}_u^0$ via (\ref{hat_bar_Delta}) with the complexity to be $\mathcal{O}(4T_3L_u^2+4L_u^3 + 4T_1L_u^2+4L_u^3 + 2L_uT_1T_3+2T_1L_u^2)$. Therefore the total computational complexity for the TDE-based scheme is
\begin{align}
   \mathcal{O}(& 8(T_3^2+T_1^2)(T_1+T_3) +  \notag \\ & 2L_u^2(10L_u+5T_3+6T_1)  + 2L_uT_1T_3 ).
\end{align}

For the EMS-based scheme, once finishing the estimation of the AoA using Algorithm~\ref{alg1}, we use Algorithm~\ref{alg3} to estimate the AoD. Firstly, we need to compute (\ref{Pphi}) for $M_A$ times, resulting in the complexity to be $\mathcal{O}( 2M_AT_1(2T_1-1) )$. Secondly, we use the bisection search to find the minimum   of $\mathcal{P}_i(\phi)$ with $\langle \log_{2}(4/(M_A\epsilon)) \rangle$ iterations, where $\epsilon$ is the predefined error of the bisection search. We compute two bisection points in each iteration, leading to the complexity to be $\mathcal{O}( 4\langle \log_{2}(4/(M_A\epsilon)) \rangle T_1(2T_1-1) )$. We also use Algorithm~\ref{alg2} to pair the AoA and AoD. Therefore the total computational complexity for the EMS-based scheme is
\begin{align}
   \mathcal{O}(& 8T_3^2(T_1+T_3) +  2L_u^2(7L_u+5T_3+3T_1) + \notag \\ & 2L_uT_1T_3  + 4\langle \log_{2}(4/(M_A\epsilon)) \rangle T_1(2T_1-1) ).
\end{align}

For the OMP-based and SWOMP-based schemes, where the AoA and AoD are first quantized into $G_r$ and $G_t$ grids, respectively, and then the AoA and AoD of $L_u$ paths are estimated using compressed sensing algorithms successively, the computational complexity is~\cite{javier2018frequency}
\begin{align}\label{GtGr}
  & \mathcal{O}( L_uKT_1T_3G_rG_t ).
\end{align}

For the DGMP-based scheme, where the mainlobe is first searched and then the compressed sensing method is used for fine-grained search, the computational complexity is
\begin{align}
  & \mathcal{O}( KT_1T_3(G_rG_t + N_AM_A) ).
\end{align}

Since $G_r>T_3$ and $G_t>T_1$, the computational complexity of the TDE-based and EMS-based schemes is much lower than that of the SWOMP-based, OMP-based and DGMP-based channel estimation schemes.


\section{Hybrid Precoding and Combining Matrices Design}
The proposed TDE-based and EMS-based channel estimation schemes can work for any $\boldsymbol{W}^k$ and $\boldsymbol{F}_u^k$ with full row rank and full column rank, respectively. However, different channel conditions with different AoA and AoD can result in different received signal power. Sometimes the channel estimation performance might be poor due to the low received signal power. Therefore, it is necessary to design $\boldsymbol{W}^k$ and $\boldsymbol{F}_u^k$ before channel estimation so that the received signal power keeps almost the same for any AoA and AoD to guarantee the robust channel estimation performance, i.e., almost the same channel estimation performance for any AoA and AoD. Since it is a very strong argument to design $\boldsymbol{W}^k$ and $\boldsymbol{F}_u^k$ with the same gain for any AoA and AoD, we design $\boldsymbol{W}^k$ and $\boldsymbol{F}_u^k$ so that the received signal power keeps almost the same instead of absolutely the same for any AoA and AoD.

According to (\ref{vecRuk}), the expectation of the received signal power neglecting the noise term can be expressed as
\begin{align}
  &\mathbb{E}\{\| \textrm{vec}(\boldsymbol{R}_u^k) \|_F^2 \} \notag \\
  &= \gamma^2 \mathbb{E} \bigg\{ \textrm{Tr} \bigg(  \left((\boldsymbol{F}_u^k)^T \otimes \boldsymbol{W}^k \right)\left(\boldsymbol{A}_{u,T}^* \circ \boldsymbol{A}_{u,R} \right) \boldsymbol{v}_u^k  (\boldsymbol{v}_u^k)^H \notag \\& ~~~~\left(\left((\boldsymbol{F}_u^k)^T \otimes \boldsymbol{W}^k \right)\left(\boldsymbol{A}_{u,T}^* \circ \boldsymbol{A}_{u,R} \right)\right)^H  \bigg) \bigg\} \notag \\
  &= \gamma^2 \textrm{Tr} \bigg(  \left((\boldsymbol{F}_u^k)^T \otimes \boldsymbol{W}^k \right)\left(\boldsymbol{A}_{u,T}^* \circ \boldsymbol{A}_{u,R} \right)  \mathbb{E} \big\{ \boldsymbol{v}_u^k  (\boldsymbol{v}_u^k)^H \big\} \notag \\& ~~~~  \left(\left((\boldsymbol{F}_u^k)^T \otimes \boldsymbol{W}^k \right)\left(\boldsymbol{A}_{u,T}^* \circ \boldsymbol{A}_{u,R} \right)\right)^H  \bigg)  \notag \\
  &= \sigma_\delta \textrm{Tr} \bigg(  \left((\boldsymbol{F}_u^k)^T \otimes \boldsymbol{W}^k \right)\left(\boldsymbol{A}_{u,T}^* \circ \boldsymbol{A}_{u,R} \right)  \notag \\& ~~~~   \left(\left((\boldsymbol{F}_u^k)^T \otimes \boldsymbol{W}^k \right)\left(\boldsymbol{A}_{u,T}^* \circ \boldsymbol{A}_{u,R} \right)\right)^H  \bigg)  \notag \\
  &\overset{(a)}= \sigma_\delta \left\| \left( (\boldsymbol{F}_u^k)^T \boldsymbol{A}_{u,T}^* \right) \circ \left( \boldsymbol{W}^k \boldsymbol{A}_{u,R} \right) \right\|_F^2 \notag \\&= \sigma_\delta \sum_{i=1}^{L_u} \left\| \left( (\boldsymbol{F}_u^k)^T \boldsymbol{\alpha}^* (M_A,\phi_{u,i}) \right) \otimes \left( \boldsymbol{W}^k \boldsymbol{\alpha} (N_A,\theta_{u,i}) \right) \right\|_2^2 \notag \\
  &\overset{(b)}= \sigma_\delta \sum_{i=1}^{L_u} \left\|  (\boldsymbol{F}_u^k)^T \boldsymbol{\alpha}^* (M_A,\phi_{u,i}) \right\|_2^2 \left\| \boldsymbol{W}^k \boldsymbol{\alpha} (N_A,\theta_{u,i}) \right\|_2^2
\end{align}
where $\sigma_\delta \boldsymbol{I}_{L_u} \triangleq \gamma^2\mathbb{E} \big\{ \boldsymbol{v}_u^k  (\boldsymbol{v}_u^k)^H \big\}$ if the gain of each channel path independently obeys the complex Gaussian distribution with zero mean and the same variance~\cite{javier2018frequency,alkhateeb2014channel}, and the equalities marked by (a) and (b) hold due to $(\boldsymbol{AB}) \circ (\boldsymbol{CD}) = (\boldsymbol{A} \otimes \boldsymbol{C})(\boldsymbol{B} \circ \boldsymbol{D})$ and $\| \boldsymbol{a} \otimes \boldsymbol{b} \|_2^2 = \|\boldsymbol{a}\|_2^2\|\boldsymbol{b}\|_2^2$, respectively. From the above discussion, the power of the received signal is the sum of that from all $L_u$ paths. Since $L_u$ channel paths are mutually independent, we require the power from each path keeps almost the same, i.e., $\|  (\boldsymbol{F}_u^k)^T \boldsymbol{\alpha}^* (M_A,\phi) \|_2^2 \| \boldsymbol{W}^k \boldsymbol{\alpha} (N_A,\theta) \|_2^2$ keeps almost the same, where $\theta_{u,i}$ and $\phi_{u,i}$ have been replaced by $\theta$ and $\phi$, respectively, to ease the notation. Note that $\theta$ and $\phi$ are independent while $\boldsymbol{F}_u^k$ and $\boldsymbol{W}^k $ are also independent. Therefore, we can separate $\|  (\boldsymbol{F}_u^k)^T \boldsymbol{\alpha}^* (M_A,\phi) \|_2^2 \| \boldsymbol{W}^k \boldsymbol{\alpha} (N_A,\theta) \|_2^2$ into two terms as $\left\|  (\boldsymbol{F}_u^k)^T \boldsymbol{\alpha}^* (M_A,\phi) \right\|_2^2$ and $\left\| \boldsymbol{W}^k \boldsymbol{\alpha} (N_A,\theta)  \right\|_2^2$. Then we require each term keeps almost the same. In the following, we focus on the design of $\boldsymbol{W}^k $ aiming at keeping $\left\| \boldsymbol{W}^k \boldsymbol{\alpha} (N_A,\theta)  \right\|_2^2$ almost the same for any $\theta \in [-1,1)$. The design of $\boldsymbol{F}_u^k$ is similar.

We consider the row-wise design of $\boldsymbol{W}^k$. By defining $\boldsymbol{w}_{n} \in{\mathbb{C}^{N_A}}$ as the $n$th column of $(\boldsymbol{W}^k)^H$, $n=1,2,\ldots,T_3$, we have
\begin{equation}\label{WT3}
  \left\| \boldsymbol{W}^k \boldsymbol{\alpha} (N_A,\theta)  \right\|_2^2 = \sum_{n=1}^{T_3} \left| \boldsymbol{w}_{n}^H \boldsymbol{\alpha} (N_A,\theta) \right|^2.
\end{equation}
We divide $\theta\in[-1,1)$ into $T_3$ nonoverlapping parts, where the $n$th part is denoted as $\Upsilon_n \triangleq [\Upsilon_n^1, \Upsilon_n^2 ) = [-1+{2(n-1)}/{T_3} , -1+{2n}/{T_3} )$ with $\Upsilon_1 \cup \Upsilon_2 \cup \ldots \cup \Upsilon_{T_3} = [-1,1)$ and $\Upsilon_m \cap \Upsilon_n = \emptyset$, $\forall m,n\in \{1,2,\ldots,T_3\}$, $m\neq{n}$. Since $\boldsymbol{w}_{n}$ only contributes to $\Upsilon_n, n=1,2,\ldots,T_3$~\cite{alkhateeb2014channel,xiao2016hi,xiao2018enhanced}, we set
\begin{equation}\label{86}
  \left| \boldsymbol{w}_{n}^H \boldsymbol{\alpha} (N_A,\theta) \right|^2 =  \left\{ \begin{array}{ll}
\xi,  & \theta  \in \Upsilon_n,   \\
0,  & \text{otherwise}.
\end{array} \right.
\end{equation}
to keep $\left\| \boldsymbol{W}^k \boldsymbol{\alpha} (N_A,\theta)  \right\|_2^2$ almost the same for any $\theta \in [-1,1)$, where $\xi$ is a constant. In fact, we may introduce a phase term $e^{j\mathcal{Q}(\theta)}$ to provide extra degree of freedom for the design of $\boldsymbol{w}_{n}$, where $\mathcal{Q}(\theta)$ is a function of $\theta$. Then we have
\begin{equation}\label{87}
  \boldsymbol{w}_{n}^H \boldsymbol{\alpha} (N_A,\theta) =  \left\{ \begin{array}{ll}
\sqrt{\xi}e^{j\mathcal{Q}(\theta)},  & \theta  \in \Upsilon_n,  \\
0,  & \text{otherwise}.
\end{array} \right.
\end{equation}
Since it is difficult to obtain $\boldsymbol{w}_{n}$ with the continuous $\theta$, we approximate the continuous $\theta$ with $M$ discrete samples inspired by~\cite{alkhateeb2014channel}, where the $m$th sample is denoted as $\theta_m \triangleq -1+2(m-1)/M$.
We define $\boldsymbol{\rho}  \in{\mathbb{C}^{M}}$ with $\boldsymbol{\rho}[m]$ denoted by
\begin{align}
  \boldsymbol{\rho}[m] = \left\{ \begin{array}{ll}
\sqrt{\xi}e^{j\mathcal{Q}(\theta_m)},  & \theta_m  \in \Upsilon_n,  \\
0,  & \text{otherwise}.
\end{array} \right.
\end{align}
Then the continuous formulation expressed in (\ref{87}) can be converted into the following discrete problem as
\begin{equation}\label{LSproblem}
  \boldsymbol{A}^H \boldsymbol{w}_{n} = \boldsymbol{\rho}
\end{equation}
where $ \boldsymbol{A} \triangleq [\boldsymbol{\alpha} (N_A,\theta_1),\boldsymbol{\alpha} (N_A,\theta_2),\ldots,\boldsymbol{\alpha} (N_A,\theta_M)] \in{\mathbb{C}^{N_A\times M}}$. The LS estimation of $\boldsymbol{w}_{n} $ from \eqref{LSproblem} is
\begin{align}\label{92}
  \hat{\boldsymbol{w}}_{n} &= \left(\boldsymbol{A}^H\right)^\dag \boldsymbol{\rho} = \left(\boldsymbol{A} \boldsymbol{A}^H\right)^{-1} \boldsymbol{A} \boldsymbol{\rho} \notag \\ &\overset{(a)}= \frac{N_A}{M}\boldsymbol{A} \boldsymbol{\rho} = \frac{N_A}{M} \sum_{m=1}^{M} \boldsymbol{\alpha} (N_A,\theta_m) \boldsymbol{\rho}[m]
\end{align}
where the equality marked by (a) holds due to $\boldsymbol{A} \boldsymbol{A}^H = M\boldsymbol{I}_{N_A}/N_A$. To guarantee the existence of $(\boldsymbol{A}^H)^\dag$, we only require $M \ge N_A$. As $M$ grows to be infinity, the sampling interval reduces to be zero, which implies that the discrete variable $\theta_m$ becomes a continuous variable $\theta$. Then (\ref{92}) can be expressed as
\begin{align}\label{93}
  \hat{\boldsymbol{w}}_{n} &= \frac{N_A}{2} \lim_{M\to\infty}  \sum_{m=1}^{M} \boldsymbol{\alpha} (N_A,\theta_m) \boldsymbol{\rho}[m] \frac{2}{M} \notag \\ &= \frac{N_A}{2} \int_{\Upsilon_n^1}^{\Upsilon_n^2}  \boldsymbol{\alpha} (N_A,\theta) \sqrt{\xi} e^{-j\mathcal{Q}(\theta)} \textrm{d}\theta.
\end{align}
To evaluate \eqref{93}, we first determine $\mathcal{Q}(\theta)$. Inspired by~\cite{chen2019beam}, we may set $\mathcal{Q}(\theta) = a\theta+b$ for simplicity, where $a$ and $b$ are two variables to be determined. In the following, we will address how to determine these two variables.

From (\ref{93}), we have
\begin{align}\label{94}
  & \hat{\boldsymbol{w}}_{n} = \frac{\sqrt{\xi N_A}}{2} \int_{\Upsilon_n^1}^{\Upsilon_n^2}  \left[1,e^{j\pi\theta},\ldots,e^{j\pi\theta(N_A-1)}\right]^{T}  e^{-j(a\theta+b)} \textrm{d}\theta  \notag \\
  &= \frac{\sqrt{\xi N_A}e^{-jb}}{2}  \bigg[ \frac{  e^{-ja  \Upsilon_n^2} - e^{-ja \Upsilon_n^1  } }{-ja}  ,   \ldots, \notag \\ &~~~~ \frac{ e^{j((N_A-1)\pi-a)  \Upsilon_n^2  } - e^{j((N_A-1)\pi-a) \Upsilon_n^1 } }{j((N_A-1)\pi-a)}  \bigg]^{T}.
\end{align}

According to (\ref{86}), all the power of $\boldsymbol{w}_{n}$ should be concentrated on $\theta  \in \Upsilon_n$, which is an ideal assumption commonly used in the existing literature, such as~\cite{chen2020two}. In practice, it is difficult to achieve (\ref{86})~\cite{alkhateeb2014channel}. Therefore, we concentrate as much power of $\hat{\boldsymbol{w}}_{n}$ as possible on $\theta  \in \Upsilon_n$.

To maximize the ratio of the power within $\theta  \in \Upsilon_n$ over the total power, we first define the power ratio as
\begin{equation}\label{95}
  \mathcal{S}(a,b)  \triangleq  \frac{\int_{\Upsilon_n^1}^{\Upsilon_n^2} \left| \hat{\boldsymbol{w}}_{n}^H \boldsymbol{\alpha} (N_A,\theta) \right|^2 \textrm{d}\theta }{\int_{-1}^{1} \left| \hat{\boldsymbol{w}}_{n}^H \boldsymbol{\alpha} (N_A,\theta) \right|^2 \textrm{d}\theta},
\end{equation}
where the numerator represents the power of $\boldsymbol{w}_{n}$ within $\theta  \in \Upsilon_n$ and the denominator represents the total power. We have $\mathcal{S}(a,b) \in[0,1]$. Then the objective is
\begin{equation}\label{PowerRatio}
  \max_{a,b} ~ \mathcal{S}(a,b).
\end{equation}

The numerator of (\ref{95}) can be simplified as
\begin{align}\label{numerator}
  &\int_{\Upsilon_n^1}^{\Upsilon_n^2} \left| \hat{\boldsymbol{w}}_{n}^H \boldsymbol{\alpha} (N_A,\theta) \right|^2 \textrm{d}\theta \notag \\ &=   \hat{\boldsymbol{w}}_{n}^H \left(\int_{\Upsilon_n^1}^{\Upsilon_n^2} \boldsymbol{\alpha} (N_A,\theta) \boldsymbol{\alpha}^H (N_A,\theta) \textrm{d}\theta \right) \hat{\boldsymbol{w}}_{n}
  \notag \\ &\triangleq    \hat{\boldsymbol{w}}_{n}^H \boldsymbol{X} \hat{\boldsymbol{w}}_{n},
\end{align}
where $\boldsymbol{X} \triangleq \int_{\Upsilon_n^1}^{\Upsilon_n^2} \boldsymbol{\alpha} (N_A,\theta) \boldsymbol{\alpha}^H (N_A,\theta)\textrm{d}\theta$. In fact, the entry at the $m$th row and $n$th column of $\boldsymbol{X}$ can be computed as
\begin{align}\label{XofNumerator}
  \boldsymbol{X}[m,n] &= \int_{\Upsilon_n^1}^{\Upsilon_n^2} \frac{1}{N_A} e^{j\pi\theta(m-1)} e^{-j\pi\theta(n-1)} \textrm{d}\theta \notag \\ &= \frac{ e^{j(m-n)\pi  \Upsilon_n^2  } - e^{j(m-n)\pi \Upsilon_n^1 } }{jN_A(m-n)\pi}.
\end{align}

The denominator of (\ref{95}) is $2\hat{\boldsymbol{w}}_{n}^H \hat{\boldsymbol{w}}_{n}/N_A$, which can be obtained by substituting $\Upsilon_n^1=-1$ and $\Upsilon_n^2=1$ into \eqref{numerator} and \eqref{XofNumerator}.

Since we require $\|\hat{\boldsymbol{w}}_{n}\|_2^2 = T_2/T_3$ to guarantee $\| \boldsymbol{W}_B^k \boldsymbol{W}_R \| _F ^2= 1$, $\mathcal{S}(a,b)$ in (\ref{95}) can be further expressed as
\begin{align}\label{98}
  &\mathcal{S}(a,b) = \frac{ N_A \hat{\boldsymbol{w}}_{n}^H \boldsymbol{X} \hat{\boldsymbol{w}}_{n} }{2 \hat{\boldsymbol{w}}_{n}^H \hat{\boldsymbol{w}}_{n}} = \frac{N_A T_3}{2 T_2} \hat{\boldsymbol{w}}_{n}^H \boldsymbol{X} \hat{\boldsymbol{w}}_{n}  \\&= \frac{N_A T_3}{2 T_2} \sum_{m=1}^{N_A}\sum_{n=1}^{N_A} \boldsymbol{X}[m,n] \hat{\boldsymbol{w}}_{n}^H[m] \hat{\boldsymbol{w}}_{n}[n] \notag \\
  &= \frac{\xi N_A T_3}{j8 T_2} \sum_{m=1}^{N_A}\sum_{n=1}^{N_A} \frac{e^{j(m-n)\pi  (\Upsilon_n^2-\Upsilon_n^1)  } - 1}{(m-n)\pi} \notag \\&~~~~ \frac{e^{-j((m-1)\pi-a)  (\Upsilon_n^2-\Upsilon_n^1)  } - 1}{(m-1)\pi-a} \frac{ e^{j((n-1)\pi-a)  (\Upsilon_n^2-\Upsilon_n^1)  } - 1 }{ (n-1)\pi-a }.\notag
\end{align}

From (\ref{98}), $\mathcal{S}(a,b)$ is only determined by $a$ and is independent of $b$. Therefore we set $b=0$ for simplicity. Then $\mathcal{S}(a,b)$ can be denoted as $\mathcal{S}(a)$ to ease the notation.

In the following, we will show that $\mathcal{S}(a)$ is a symmetric function with symcenter to be $(N_A-1)\pi/2$. Based on \eqref{98}, we have
\begin{align}\label{100}
  &\mathcal{S}((N_A-1)\pi-a) \\
  &=  \frac{\xi N_A T_3}{j8 T_2} \sum_{m=1}^{N_A}\sum_{n=1}^{N_A} \frac{e^{j(m-n)\pi  (\Upsilon_n^2-\Upsilon_n^1)  } - 1}{(m-n)\pi} \notag \\&~~~~ \frac{e^{-j((m-N_A)\pi+a)  (\Upsilon_n^2-\Upsilon_n^1)  } - 1}{(m-N_A)\pi+a} \frac{ e^{j((n-N_A)\pi+a)  (\Upsilon_n^2-\Upsilon_n^1)  } - 1 }{ (n-N_A)\pi+a } \notag \\
  &\overset{(a)}= \frac{\xi N_A T_3}{j8 T_2} \sum_{\bar{m}=1}^{N_A}\sum_{\bar{n}=1}^{N_A} \frac{e^{j(\bar{m}-\bar{n})\pi  (\Upsilon_n^2-\Upsilon_n^1)  } - 1}{(\bar{m}-\bar{n})\pi} \notag \\&~~~~ \frac{e^{-j((\bar{m}-1)\pi-a)  (\Upsilon_n^2-\Upsilon_n^1)  } - 1}{(\bar{m}-1)\pi-a} \frac{ e^{j((\bar{n}-1)\pi-a)  (\Upsilon_n^2-\Upsilon_n^1)  } - 1 }{ (\bar{n}-1)\pi-a } \notag
\end{align}
where the equality marked by (a) holds by defining $\bar{m} \triangleq N_A-n+1$ and $\bar{n} \triangleq N_A-m+1$. Comparing (\ref{98}) and (\ref{100}), we have $\mathcal{S}(a) = \mathcal{S}((N_A-1)\pi-a)$, indicating that $\mathcal{S}(a)$ is a symmetric function with symcenter to be $(N_A-1)\pi/2$. Therefore, to find the maximum of $\mathcal{S}(a)$ as in \eqref{PowerRatio}, we only need to search $a\ge (N_A-1)\pi/2$, which can reduce the searching complexity by half according to the symmetry of $\mathcal{S}(a)$.

Now we determine an upper bound for the search of optimal $a$. Without an upper bound, we have to search $a$ from $(N_A-1)\pi/2$ to infinity, which is computationally intractable.
Term $(e^{-j((m-1)\pi-a)  (\Upsilon_n^2-\Upsilon_n^1)  } - 1)(e^{j((n-1)\pi-a)  (\Upsilon_n^2-\Upsilon_n^1)  } - 1)$ in the numerator of (\ref{98}) is a periodic function of $a$ with the period as $2\pi/(\Upsilon_n^2-\Upsilon_n^1)=T_3\pi$. In the denominator of (\ref{98}), $((m-1)\pi-a)((n-1)\pi-a)$ is a quadratic function of $a$, which is monotonically increasing and always greater than zero when $a>(N_A-1)\pi$. Since the monotonically increasing denominator leads to the decrease of the ratio, also considering the variation of the numerator of (\ref{98}) within the period of $T_3\pi$, an upper bound for the search of optimal $a$ is $(N_A-1)\pi+T_3\pi$. Therefore, we can search an optimal $a$ from (\ref{98}) in the range of $a\in\big[(N_A-1)\pi/2, (N_A-1)\pi+T_3\pi\big]$.

It is difficult to search the optimal $a$ from (\ref{98}) based on the existing fast search algorithms as $\mathcal{S}(a)$ is nonmonotonic and varying with $a$. We sample $\mathcal{S}(a)$ in $[(N_A-1)\pi/2, (N_A-1)\pi+T_3\pi]$ with $N$ equally spaced points, where the $n$th point is denoted as $a_n \triangleq (N_A-1)\pi/2+((N_A-1)\pi/2+T_3\pi)(n-1)/N$. We find the optimal $a_n$ with the largest $\mathcal{S}(a_n)$ via
\begin{equation}\label{99}
  \hat{a} = \arg \underset{n=1,2,\ldots,N} {\max} ~ \mathcal{S}(a_n).
\end{equation}
After obtaining $a$ and setting $b=0$, we can determine $\xi$ in \eqref{94} by normalizing $\hat{\boldsymbol{w}}_{n}$ as $\|\hat{\boldsymbol{w}}_{n}\|_2^2 =T_2/T_3 $, which guarantees $\| \boldsymbol{W}_B^k \boldsymbol{W}_R \| _F ^2= 1$. Then $\hat{\boldsymbol{w}}_{n}$ is finally obtained.

In summary, to design a hybrid combining matrix, we first obtain the optimal $a$ from (\ref{99}) and set $b=0$. Then we obtain $\boldsymbol{w}_{n}$ from (\ref{94}). Finally $\boldsymbol{W}^k$ can be denoted as $\boldsymbol{W}^k=[\boldsymbol{w}_{1},\boldsymbol{w}_{2},\ldots,\boldsymbol{w}_{T_3}]^H$. The design of $\boldsymbol{F}_u^k$ is similar to that of $\boldsymbol{W}^k $.

\section{Simulation Results}
Now we evaluate the performance of the proposed TDE-based and EMS-based schemes. We consider a multi-user mmWave massive MIMO system, where the BS equipped with $N_A=64$ antennas and $N_R=4$ RF chains serves $U=4$ users each equipped with $M_A=16$ antennas and $M_R=1$ RF chain. The number of resolvable paths in mmWave channel is set to be $L_{u}=3$ with $g_{u,i}\thicksim\mathcal{CN}(0,1)$ for $i=1,2,\ldots,L_u$~\cite{javier2018frequency,alkhateeb2014channel}. The delay of each channel path denoted as $\tau_{u,i}$ obeys the uniform distribution $[0,5T_s]$, which means the delay spread can be 5 samples at most. As in~\cite{javier2018frequency}, we use $K=16$ OFDM subcarriers for the pilot transmission in frequency-selective mmWave channels. The number of delay taps of the channel is set to be $D=4$. We set the predefined threshold $\epsilon=0.001$ for Algorithm~\ref{alg3} and  $N=1000$ for \eqref{99}. For the SWOMP-based scheme~\cite{javier2018frequency}, the OMP-based scheme~\cite{venugopal2017time} and the DGMP-based scheme~\cite{dai2016estimation}, we set $G_r=G_t=90$ for \eqref{GtGr} according to~\cite{javier2018frequency}.

\begin{figure}[!t]
\centering
\includegraphics[width=84mm]{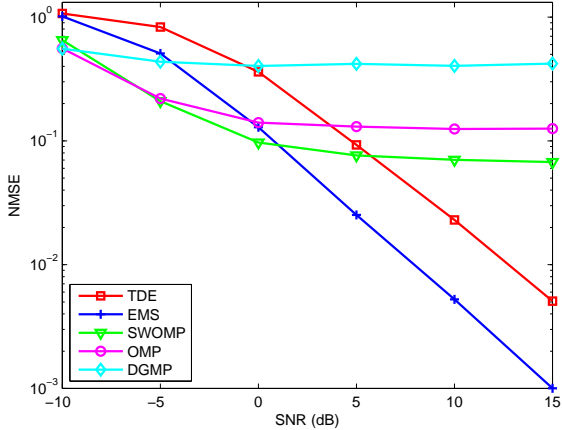}
\caption{Comparisons of channel estimation performance for different schemes in terms of SNR.}
\label{Fig2}
\end{figure}

\begin{figure}[!t]
\centering
\includegraphics[width=84mm]{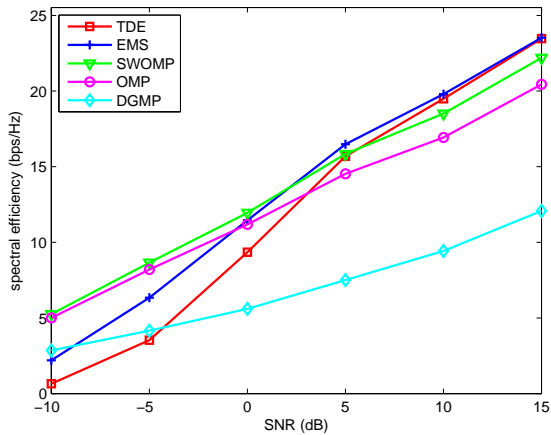}
\caption{Comparisons of spectral efficiency for different schemes in terms of SNR.}
\label{Fig3}
\end{figure}

\begin{figure}[!t]
\centering
\includegraphics[width=84mm]{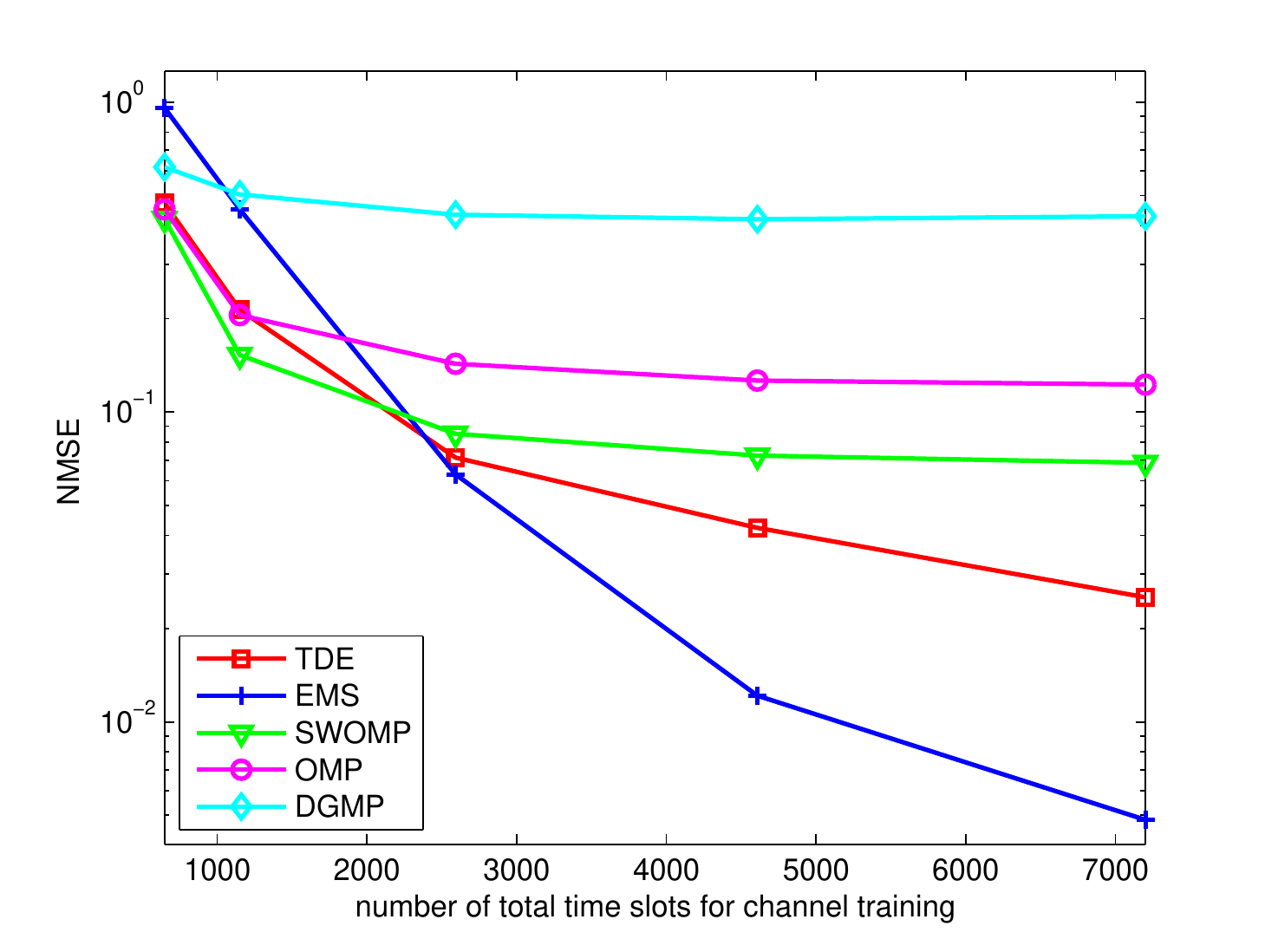}
\caption{Comparisons of channel estimation performance for different schemes in terms of the number of total time slots for channel training.}
\label{FIG6}
\end{figure}

\begin{figure}[!t]
\centering
\includegraphics[width=84mm]{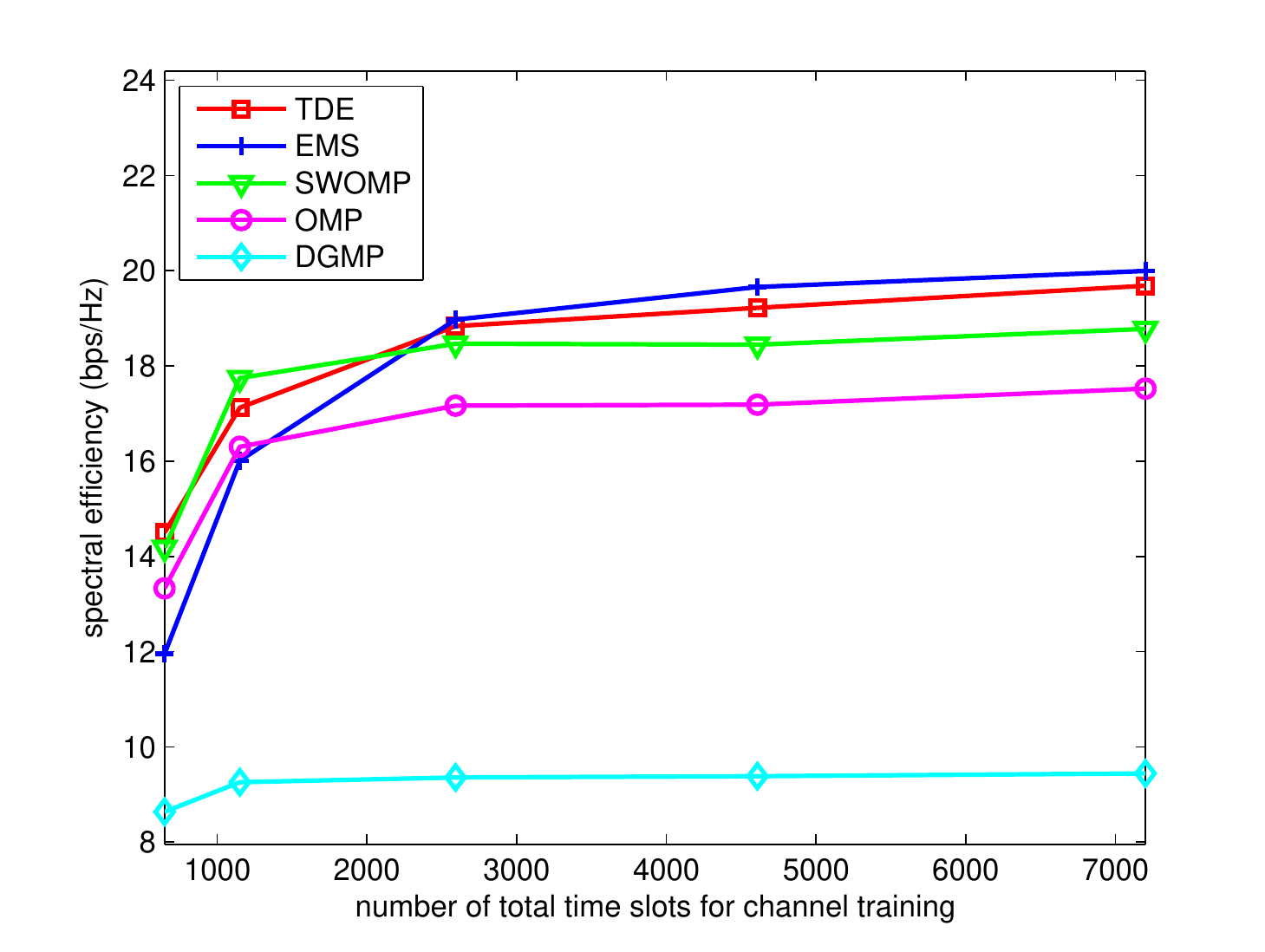}
\caption{Comparisons of spectral efficiency for different schemes in terms of the number of total time slots for channel training.}
\label{FIG7}
\end{figure}

As shown in Fig.~\ref{Fig2}, we compare the channel estimation performance for the TDE-based scheme, the EMS-based scheme, the SWOMP-based scheme, the OMP-based scheme and the DGMP-based scheme in terms of different SNRs. The channel estimation performance is measured by normalized mean-squared error (NMSE), which is defined as
\begin{equation}
  \textrm{NMSE} \triangleq \frac{ \sum_{u=1}^{U} \sum_{k=0}^{K-1} \| \hat{\boldsymbol{H}}_{u}^k - \boldsymbol{H}_{u}^k \|_F^2 }{ \sum_{u=1}^{U} \sum_{k=0}^{K-1} \| \boldsymbol{H}_{u}^k \|_F^2 }.
\end{equation}
We set $T_1=12$ and $T_2=8$. Then $T_3=T_2N_R=32$. For the TDE-based scheme, the number of total time slots for pilot training is $(K+2)UT_1T_2=6,912$, which is much smaller than the maximum limitation of time slots for pilot training 400,000 discussed in Section II.C.  To make fair comparison, we fix the total time slots for pilot training to be  $6,912$ for the DGMP-based scheme, SWOMP-based scheme and OMP-based scheme as well as the EMS-based scheme. Due to the fact that the SVD operation is sensitive to the noise at low SNR region, the TDE-based and EMS-based schemes perform worse than the existing schemes. However, at high SNR region, both the TDE-based and the EMS-based schemes are much better than the existing schemes. At SNR of 10 dB, the TDE-based scheme has 67.3\%, 81.5\%, and 94.3\% performance improvement compared with the SWOMP-based, OMP-based and DGMP-based schemes, respectively, while the EMS-based scheme has 92.5\%, 95.8\%, and 98.7\% performance improvement compared with the SWOMP-based, OMP-based and DGMP-based schemes, respectively. The reason for the unsatisfactory performance is that the SWOMP-based and OMP-based schemes ignore the power leakage due to the limited beamspace resolution and the DGMP-based scheme only estimates a single path while our proposed TDE-based and EMS-based schemes can simultaneously estimate multipaths. Note that the proposed TDE-based and EMS-based schemes are not impaired by the power leakage and can achieve high resolution.

In Fig.~\ref{Fig3}, we compare the spectral efficiency for the proposed TDE-based scheme, the EMS-based scheme, the SWOMP-based scheme, the OMP-based scheme and the DGMP-based scheme in terms of SNR. From the figure, the proposed schemes achieve better performance than the others at the high SNR region. At SNR 10 dB, the TDE-based scheme has 5.3\%, 15.1\%, and 106.7\% performance improvement compared with the SWOMP-based, OMP-based and DGMP-based schemes, respectively, while the EMS-based scheme has 6.9\%, 16.9\%, and 110.0\% performance improvement compared with the SWOMP-based, OMP-based and DGMP-based schemes, respectively. The reason for the smaller spectral efficiency gap between different schemes than the NMSE gap is that the NMSE performance is much more sensitive to the AoA and AoD accuracy, while the spectral efficiency performance is determined by the beamforming gain and is less sensitive to the AoA and AoD accuracy.

In Fig.~\ref{FIG6}, we compare the channel estimation performance for different schemes in terms of the number of total time slots for channel training. We fix SNR to be 10 dB. For the TDE-based scheme, the number of total time slots for pilot training is $(K+2)UT_1T_2=72T_1T_2$, which is fairly set the same for the DGMP-based scheme, SWOMP-based scheme and OMP-based scheme as well as the EMS-based scheme. For simplicity, we set $T_1=T_2$ and use different $T_1$ in the simulation. From the figure, when the number of total time slots for pilot training is large, the proposed schemes achieve better performance than the other schemes. Fixing the number of total time slots to be 7,200, which corresponds to $T_1=T_2=10$, the TDE-based scheme has 63.2\%, 79.4\%, and 94.1\% improvement compared with the SWOMP-based, OMP-based and DGMP-based schemes, respectively, while the EMS-based scheme has 93.0\%, 96.1\%, and 98.9\% improvement compared with the SWOMP-based, OMP-based and DGMP-based schemes, respectively.

In Fig.~\ref{FIG7}, we compare the spectral efficiency for different schemes in terms of the number of total time slots for channel training. The parameters for the simulation are set the same as those for Fig.~\ref{FIG6}. We observe that the TDE-based and EMS-based schemes can achieve better performance than the SWOMP-based, OMP-based and DGMP-based schemes when the number of total time slots for channel training is more than 3,000. When the number of total time slots for channel training is more than 4,608, which corresponds to $T_1=T_2=8$, the spectral efficiency of the TDE-based and EMS-based schemes keeps almost the same, indicating that $T_1=T_2=8$ is enough to obtain the full channel station information and thus can achieve the maximal spectral efficiency.

\section{Conclusions}
In this paper, we have proposed two high-resolution channel estimation schemes, i.e., the TDE-based scheme and EMS-based scheme. Following these two schemes, we have also developed a hybrid precoding and combining matrices design method so that the received signal power keeps almost the same for any AoA and AoD to guarantee robust channel estimation performance. Additionally, we have also compared the proposed two schemes with the existing channel estimation schemes in terms of computational complexity. Simulation results have verified the effectiveness of our work and have shown that the proposed schemes outperform the existing schemes. Future work will focus on  the high-resolution channel estimation for frequency-selective mmWave massive MIMO systems equipped with other forms of uniform arrays, such as the uniform planar arrays (UPAs).

\bibliographystyle{IEEEtran}
\bibliography{IEEEabrv,IEEEexample}

\end{document}